\documentclass[12pt]{article}
\DeclareMathAlphabet{\scr}{U}{rsfs}{m}{n}
\usepackage{latexsym}
\usepackage[mathscr]{eucal}
\usepackage{amsfonts}
\usepackage{amscd}
\usepackage{cite}
\usepackage{amssymb}
\usepackage[centertags]{amsmath}
\usepackage{dsfont}
\usepackage{yfonts} 
\usepackage{enumerate}
\usepackage[center]{subfigure}
\usepackage[vcentermath]{youngtab}

\newcommand{\cleqn}{\setcounter{equation}{0}}
\newcommand{\newc}{\newcommand}
\newc{\eps}{\epsilon}
\newc{\lam}{\lambda}
\newc{\Lam}{\Lambda}
\newc{\ra}{\rightarrow}
\newc{\lra}{\leftrightarrow}
\newc{\wtilde}{\widetilde}
\newc{\ie}{{\it i.e.}}
\newc{\eg}{{\it e.g.}}
\newc{\rpv}{\not\!\! M_p}
\newc{\lsim}{\stackrel{<}{\sim}}
\newc{\beq}{\begin{equation}}
\newc{\eeq}{\end{equation}}
\newc{\beqn}{\begin{eqnarray}}
\newc{\eeqn}{\end{eqnarray}}
\newc{\PLB}{\emph{Phys.Lett.}{\bf{B}}}
\newc{\NPB}{\emph{Nucl.Phys.}{\bf{B}}}
\newc{\mcal}{\mathcal}
\newc{\bsym}{\boldsymbol}
\newc{\nonum}{\nonumber}
\newc{\ol}{\overline}
\newc{\wt}{\widetilde}
\newc{\bs}{\boldsymbol}
\newc{\m}{\mathcal}
\begin{document}

\title{\hfill ~\\[-30mm]
       \hfill\mbox{\small UFIFT-HEP-08-7}\\[30mm]
       \textbf{Anomaly Conditions for \\[2mm] Non-Abelian Finite Family
         Symmetries }} 
\date{}
\author{\\Christoph Luhn,\footnote{E-mail: {\tt luhn@phys.ufl.edu}}~~
        Pierre Ramond\footnote{E-mail: {\tt ramond@phys.ufl.edu}}\\ \\
  \emph{\small{}Institute for Fundamental Theory, Department of Physics,}\\
  \emph{\small University of Florida, Gainesville, FL 32611, USA}}

\maketitle

\begin{abstract}
\noindent Assuming that finite family symmetries are gauged, we
derive discrete anomaly conditions for various non-Abelian groups.
We thus provide new constraints for flavor model building, in which discrete
non-Abelian symmetries are employed to explain the tri-bimaximal mixing
pattern  in the lepton sector. 
\end{abstract}
\thispagestyle{empty}
\vfill
\newpage
\setcounter{page}{1}

\section{Introduction}
\cleqn
Forty years after the setup of the Homestake experiment
\cite{Cleveland:1998nv}, the concept of neutrino oscillations is well
established and generally accepted to explain the solar neutrino puzzle. Solar
\cite{Ahmed:2003kj,Smy:2003jf}, atmospheric \cite{Ashie:2005ik}, reactor
\cite{Araki:2004mb}, and accelerator \cite{Ahn:2002up} neutrino 
experiments all made important contributions to our current knowledge of
the neutrino sector.
In particular, one finds that the measured Maki-Nakagawa-Sakata-Pontecorvo
(MNSP) leptonic mixing matrix approximately displays the so-called tri-bimaximal
pattern \cite{Harrison:2002er,Harrison:2002kp}. This fact has triggered an
overwhelming interest in non-Abelian finite groups as means to explain the
family structure of leptons and quarks.

Although differing in their details, all proposed models augment the Standard
Model (SM) gauge symmetry with a discrete symmetry. If this discrete
symmetry~$\m G$ originates in a continuous gauge symmetry~$G$ which is
spontaneously broken, one refers to it as a {\it discrete gauge symmetry}. The
assumption of a gauge origin has the advantage that the remnant discrete
symmetry $\m G$ is protected against violations by quantum gravity
effects~\cite{Krauss:1988zc}. We therefore require an underlying gauge
symmetry of the form
$$
\mathrm{SM} ~ \times ~ G \ .
$$
Whenever a new gauge symmetry is added to the SM gauge group, it
is necessary to verify that such an extension is anomaly free. With the above
symmetry structure, three new types of anomalies arise: 
\beq
\mathrm{SM} - \mathrm{SM} - G \ , \qquad 
\mathrm{SM} - G - G \ , \qquad
G - G - G \ . \nonumber
\eeq
The requirement of anomaly freedom at the level of the continuous gauge
symmetry~$G$ results in the so-called {\it discrete anomaly conditions} after
its breaking to the discrete symmetry~$\m G \subset G$.

Ib\'a\~nez and Ross~\cite{Ibanez:1991hv,Ibanez:1991pr} were the first to carry
out a systematic study of these discrete anomaly conditions in the case of~$G
= U(1)$ breaking down to~$\m G = \m Z_N$. There the potential anomalies are 
\beqn
\begin{array}{lll}\notag
SU(3)_C-SU(3)_C-U(1) \, , ~ & SU(2)_W-SU(2)_W-U(1) \, ,~ & \mathrm{Gravity}-\mathrm{Gravity}-U(1) \,
, \\[4mm]
U(1)_Y-U(1)_Y-U(1) \, , ~ & U(1)_Y-U(1)-U(1) \, , ~ & U(1)-U(1)-U(1) \, ,
\end{array}
\eeqn 
where $SU(3)_C$, $SU(2)_W$, and $U(1)_Y$ are the Standard Model gauge
groups. The investigation of their 
discrete versions revealed that the anomalies of the first row severely
constrain the allowed anomaly-free discrete gauge symmetries. Under the
assumptions that the  
light fermions of the theory, i.e those particles which do not acquire a mass
when the $U(1)$ breaks down to $\m Z_N$, are solely the Standard Model
particles, only a finite number of non-equivalent $\m Z_N$ symmetries
is possible~\cite{Dreiner:2005rd}. Adding three right-handed neutrinos to the
light particle content, thus requiring pure Dirac neutrinos, allows an
infinity of anomaly-free discrete gauge symmetries~\cite{Luhn:2007gq}.

An analogous systematic study of the discrete anomaly conditions for the case
where~$\m G$ is a non-Abelian finite group is still lacking. It is the purpose
of this article to fill this gap and provide useful constraints on flavor
models applying a non-Abelian discrete symmetry. 

Since there are only three
chiral families in Nature, any candidate finite family group~$\m G$ should
have two- or three-dimensional irreducible representations. This limits the
possibilities to finite subgroups of $SU(3)$, $SU(2)$, and $SO(3) \approx
SU(2)/\m Z_2$. Here we restrict ourselves to the groups 
$\m{PSL}_2(7)$~\cite{Luhn:2007yr}, 
$\m Z_7 \rtimes \m Z_3$~\cite{Luhn:2007sy}, 
$\Delta(27)$~\cite{deMedeirosVarzielas:2005qg,deMedeirosVarzielas:2006fc,Ma:2006ip,Ma:2007wu,Plentinger:2008nv},  
$\m S_4$~\cite{Hagedorn:2006ug,Koide:2007sr,Lam:2008rs},
$\m
A_4$~\cite{Ma:2001dn,Babu:2002dz,Babu:2005se,Zee:2005ut,Altarelli:2005yp,Altarelli:2005yx,King:2006np,Morisi:2007ft,Hirsch:2007kh,Bazzocchi:2007au,Honda:2008rs,Altarelli:2008bg,Lin:2008aj}, 
$\m D_5$~\cite{Hagedorn:2006ir,Blum:2007jz}, and  
$\m
S_3$~\cite{Pakvasa:1977in,Caravaglios:2005gw,Mohapatra:2006pu,Jora:2006dh,Koide:2006vs,Babu:2007zm,Feruglio:2007hi}.
The underlying family gauge symmetry $G_f$ must be non-Abelian, 
the natural candidates being
$$G_f ~~=~~ SU(3)\,,~ SU(2)\,,~ SO(3)\ .$$
Then, the potential anomalies are\footnote{Upon completion of this work we
  became aware of Ref.~\cite{Araki:2008ek} which focuses on discrete
  anomalies of the type $\m G - \mathrm{SM} - \mathrm{SM}$,
  see also~\cite{Babu:2004tn}. Notice that such 
  an anomaly does not exist in our approach because there is no $G_f -
  \mathrm{SM} -\mathrm{SM}$ anomaly to begin with.}
\beq 
G_f-G_f-U(1)_Y \ , \qquad  G_f - G_f - G_f \ ,\notag
\eeq
where the cubic anomaly is absent for $G_f=SU(2) ~\mathrm{or}~ SO(3)$.
In order to formulate the discrete anomaly conditions, we first need to
understand how the irreps ${\bs \rho}$ of $G_f$ decompose into irreps ${\bf
r_i}$ of the finite subgroup $\m G$. As this decomposition depends
on the finite group and its underlying gauge group, one has to discuss each
case separately. Furthermore, since the quadratic and
the cubic indices of the irreps of~$G_f$ enter the original anomaly
conditions, it is necessary to introduce the concept of {\it discrete indices}
for irreps of~$\m G$. The following table summarizes our conventions and
notation before and after the breaking of~$G_f$ into~$\m G$.

\begin{center}
\begin{tabular}{|l|c|c|c|c|} \hline&&&&\\[-4mm]
& Family Group & Irreps & Index & Constraints \\[1mm]\hline&&&&\\[-4mm]
High-Energy & $G_f$ & ${\bs \rho}$ & $I({\bs \rho})$ 
& anomaly conditions \\[1mm] \hline&&&&\\[-4mm]
Low-Energy & $\m G$ & ${\bf r_i}$ & $\wt I({\bf r_i})$ 
& discrete anomaly conditions \\[1mm] \hline
\end{tabular}
\end{center}

\vspace{2mm}

In Section~\ref{general}, we present the general method of consistently defining
discrete indices for irreps of~$\m G$ assuming that the
underlying family symmetry is $SU(3)$. We discuss the symmetries
$\m{PSL}_2(7)$ and $\m Z_7 \rtimes \m Z_3$ explicitly in
Section~\ref{pslsect}. The corresponding results for $\m G =\Delta(27),\m
S_4,\m A_4,\m D_5,\m S_3$ are shown in Section~\ref{othersect}.  
In Section~\ref{so3sect} we consider the possibility that $\m G$ originates
from $SO(3)$, leading to exactly the same discrete quadratic indices as for
$SU(3)$. Having defined the discrete indices, we determine the
discrete anomaly conditions in Section~\ref{DACsect}. We apply our discrete
anomaly conditions to existing examples of flavor models in
Section~\ref{casessect} and conclude in
Section~\ref{conclsect}. Appendices~\ref{young} and~\ref{f-proof} supplement
the proof of defining discrete indices consistently. 

Finally, for the sake of
quick reference for model builders, we summarize our discrete anomaly conditions
together with the relevant discrete indices in Appendix~\ref{mainresults}.

\section{\label{general}The Indices of  Finite Subgroups of $\bs{SU(3)}$}
\cleqn
It is the purpose of this section to provide a definition of indices for
finite subgroups $\m G$ of $SU(3)$. The underlying idea is that the finite
group should have a gauge origin. As the continuous gauge
symmetry $SU(3)$ is broken, its representations ${\bs \rho}$ decompose into a
sum of representations ${\bf r_i}$ of $\m G$. We will show that one can
consistently introduce {\it discrete indices} 
for the irreps~${\bf r_i}$ of $\m G$. These discrete
indices can be understood as the vestige of the $SU(3)$ gauge theory which is
supposed to be anomaly free. Therefore, they allow an extension of the
well-known discrete anomaly conditions for Abelian symmetries
\cite{Ibanez:1991hv,Ibanez:1991pr,Dreiner:2005rd,Luhn:2007gq} to the
non-Abelian case. 

As a starting point, let us briefly recapitulate how the indices of $SU(3)$
irreps are defined. The algebra of $SU(3)$ is given by the Gell-Mann matrices
$\lambda_a$, $a=1,...,8$. The $3 \times 3$ matrices satisfy
\cite{GellMann:1961ky,GellMann:1964xy} 
\beq
\lambda_a \, \lambda_b ~=~ \frac{2 \delta_{ab}}{3} \, \mathds{1} ~+~ i \,
f_{abc} \, \lambda_c ~+~ d_{abc} \, \lambda_c \ ,\notag
\eeq
where $\delta_{ab}$ denotes the Kronecker symbol, $f_{abc}$ are the
(antisymmetric) structure constants and $d_{abc}$ the (symmetric)
$d$-coefficients of $SU(3)$. 
Denoting the generators of the representation ${\bs \rho}$ of $SU(3)$ by
$T^{[{\bs \rho}]}_a$, we can determine the following traces
\beqn
\mathrm{Trace} \left( \left\{ \, T^{[{\bs \rho}]}_a \,,\,T^{[{\bs \rho}]}_b \, \right\}\right) &=&  \ell({\bs \rho}) \, \delta_{ab} \ , \\ 
\mathrm{Trace} \left( \left\{ \, T^{[{\bs \rho}]}_a \,,\,T^{[{\bs \rho}]}_b \, \right\}  \,T^{[{\bs \rho}]}_c \right) &=& A({\bs \rho})\, \frac{d_{abc}}{2} \ ,
\eeqn
with $\{~,~\}$ being the anticommutator. $\ell ({\bs \rho})$ and $A({\bs
  \rho})$ are respectively the quadratic and the cubic index of ${\bs
  \rho}$. These two indices correspond to the two fundamental Casimir
operators of $SU(3)$. Applying the normalization in which the generators of the
irrep ${\bs \rho}={\bf 3}$ are given by $T^{[{\bf 3}]}_a =
\lambda_a/2$ we obtain  
$$
\ell({\bf 3})~=~ \ell(\ol{\bf  3})~=~1\ , ~\qquad ~
A({\bf 3})~=~ -\, A(\ol{\bf  3})~=~1\ ,
$$
for the fundamental irrep~${\bf 3}$ and its complex conjugate~$\ol{\bf 3}$. 
For higher irreps ${\bs \rho}$ of $SU(3)$, the indices $I({\bs \rho}) =
\ell({\bs \rho}),A({\bs \rho})$  can be calculated recursively from  
the composition relation \cite{Banks:1976yg,Slansky:1981yr,Okubo:1982dt}
\beq
I({\bs \rho} \otimes {\bs \sigma}) ~=~ d({\bs \rho}) \, I({\bs \sigma}) ~+~ I({\bs \rho}) \, d({\bs \sigma}) \ ,
\label{product}
\eeq
where $d({\bs \rho})$ is the dimension of ${\bs \rho}$. For the irreps
up to dimension 27 one finds the following values.\\
{\small
\begin{center}
\begin{tabular}{|rr|r|r|}
\hline ~ &&& \\[-3mm]
\multicolumn{2}{|c|}{Irreps ${\bs \rho}$ of $SU(3)$} & $\ell({\bs \rho})$ &  $A({\bs \rho})$ \\[-3mm]~&&& \\ \hline
~&&&\\[-3mm]
~~~~~(10)\,: & {\bf 3}~~~~~       &  1~   &      1~ \\
~~~~~(01)\,: & $\ol{\bf 3}$~~~~~  &  1~   &   $-1$~ \\
~~~~~(20)\,: & {\bf 6}~~~~~       &  5~   &      7~ \\
~~~~~(02)\,: & $\ol{\bf 6}$~~~~~  &  5~   &   $-7$~  \\
~~~~~(11)\,: & {\bf 8}~~~~~       &  6~   &      0~ \\
~~~~~(30)\,: & {\bf 10}~~~~~     & 15~   &     27~ \\
~~~~~(21)\,: & {\bf 15}~~~~~      & 20~   &     14~ \\
~~~~~(40)\,: & ${\bf 15}'\hspace{-1mm}$~~~~~   & 35~   &     77~ \\
~~~~~(05)\,: & {\bf 21}~~~~~      & 70~   & $-182$~ \\
~~~~~(13)\,: & {\bf 24}~~~~~      & 50~   &  $-64$~ \\
~~~~~(22)\,: & {\bf 27}~~~~~      & 54~   &      0~ \\[1mm]\hline
\end{tabular}
\end{center}
}
\vspace{6mm}

When $SU(3)$ breaks down to the finite subgroup $\m G$, the irreps ${\bs \rho}$
decomposes into irreps~${\bf r_i}$ of the finite subgroup with multiplicities
$a_i$. 
We have
\beq
{\bs \rho} ~=~ \bigoplus_i ~a_i\, {\bf r_i} \ ,
\eeq
where the sum is over all irreps of the finite group. Since this breaking
process must be consistent with the Kronecker products of $\m G$, the 
irreps~${\bf r_i}$ inherit {\it discrete indices} 
$\wt I({\bf r_i}) = \wt\ell({\bf r_i}),\wt A({\bf r_i})$
from their parent irreps. Assuming that these discrete indices $\wt I({\bf
  r_i})$ are well-defined, we introduce the quantity 
\beq
\mathfrak I({\bs \rho})~=~a_i \, \wt I({\bf r_i}) \ ,
\eeq
to show that 
\beq
I({\bs \rho})~=~ \mathfrak I({\bs \rho}) ~\mathrm{mod~}N_I \ ,\label{consistent}
\eeq
holds true for all irreps ${\bs \rho}$. The integer $N_I$ depends only on the
type of index (quadratic or cubic) and the finite subgroup~$\m G$. Before
proving Eq.~(\ref{consistent}) for individual finite groups~$\m G$, we outline
the general procedure.

Evaluation of Eq.~(\ref{consistent}) for the
smallest $SU(3)$ irreps can be used to ``guess'' the discrete indices $\wt
I({\bf r_i})$ and the value of $N_I$. Once these numbers are given, it is
possible to prove by induction that Eq.~(\ref{consistent}) is valid for all
higher irreps of $SU(3)$ as well.\footnote{We thank Dr. Yuji Tachikawa
for his inductive proof for the discrete cubic
index of $\m{PSL}_2(7)$.}   
Since these higher irreps can be obtained by successive
multiplication with smaller irreps, the inductive step consists in showing the
validity of Eq.~(\ref{consistent}) for the product ${\bs \rho} \otimes {\bs \sigma}$,
where ${\bs \sigma}$ is an $SU(3)$ irrep which decomposes as
\beq
{\bs \sigma} ~=~ \bigoplus_i ~b_i\, {\bf r_i} \ .
\eeq
It is argued in Appendix~\ref{young} that we need only consider ${\bs \sigma} = {\bf
3},\ol{\bf 3}$ to prove our proposition. At this stage, however, we keep our
presentation general. 

\begin{itemize}
\item For ${\bs \rho} \otimes {\bs \sigma}$, the left-hand side of Eq.~(\ref{consistent})
  is obtained from Eq.~(\ref{product}), yielding
\beqn
I({\bs \rho} \otimes {\bs \sigma}) &=& a_i \, d({\bf r_i}) \, I({\bs \sigma}) 
~+~  \big( \mathfrak I({\bs \rho})~\mathrm{mod~}N_I \big) \, d({\bs \sigma}) \notag \\
&=& a_i  \underbrace{\big[ d({\bf r_i}) \, I({\bs \sigma}) \,+\,  \wt I({\bf r_i})\, d({\bs \sigma})
  \big]}_{\equiv  f_I^{\,i}({\bs \sigma})}  ~~
\mathrm{mod~~}d({\bs \sigma}) N_I \ ,
\eeqn
with $d({\bf r_i})$ denoting the dimension of the irrep ${\bf r_i}$. Notice that we have
assumed Eq.~(\ref{consistent}) for the irrep ${\bs \rho}$ in the first
step. Variation of ${\bs \rho}$ in this equation changes the parameters $a_i$
whereas the factors $f_I^{\,i}({\bs \sigma})$ remain unaffected.  

\item Next we consider the right-hand side of Eq.~(\ref{consistent}) for ${\bs \rho}
\otimes {\bs \sigma}$. This representation decomposes into the irreps of the finite
group as 
\beq
{\bs \rho} \otimes {\bs \sigma} ~=~ a_i \,  b_j ~  {\bf r_i} \otimes {\bf r_j} ~=~ a_i \, b_j \,
K^{ij}_k ~ {\bf r_k} \ .\label{decKro}
\eeq  
$K^{ij}_k$ are the multiplicities of the irrep ${\bf r_k}$ in the Kronecker
product ${\bf r_i} \otimes {\bf r_j}$. We get
\beq
\mathfrak I({\bs \rho} \otimes {\bs \sigma}) ~=~ a_i\, b_j \, K^{ij}_k \, \wt I({\bf r_k}) 
~=~  a_i \underbrace{\big[ \wt I({\bf r_i} \otimes b_j {\bf r_j})
  \big]}_{\equiv \mathfrak f_I^{\,i}({\bs \sigma})} \ ,
\eeq
where, in the last step, we require $\wt I(\cdots)$ to be linear in its
argument. Again the factors $\mathfrak f_I^{\,i}({\bs \sigma})$ depend only on
${\bs \sigma}$ and $i$ but not on ${\bs \rho}$.
\end{itemize}
With the above remarks, the proof of Eq.~(\ref{consistent}) boils down to
showing that 
\beq
f_I^{\,i}({\bs \sigma}) ~=~ \mathfrak f_I^{\,i}({\bs \sigma})~\mathrm{mod~}N_I \ .\label{effs}
\eeq
In the following sections we will discuss various finite subgroups of
$SU(3)$, presenting the decomposition of the smallest $SU(3)$ irreps, listing
the ``guessed'' values for $N_I$ and the discrete indices, and finally proving
that these definitions satisfy Eq.~(\ref{effs}) for ${\bs \sigma}={\bf 3}, \ol{\bf
  3}$. Thus the concept of discrete indices is shown to be consistent.

\section{\label{pslsect}Indices of  $\bs{\m{PSL}_2(7)}$ and $\bs{\m Z_7
    \rtimes \m Z_3}$}
\cleqn
\subsection{The Group $\bs{\m{PSL}_2(7)}$}

As our first example, we discuss the case of $\m{PSL}_2(7)$ which is the
unique simple subgroup of $SU(3)$ with complex three-dimensional irreps. 
Including the singlet, it has six irreps ${\bf r_i}$ 
$$
{\bf r_0}={\bf 1}, \quad {\bf r_1}={\bf 3}, \quad
{\bf r_2}=\ol{\bf 3}, \quad  {\bf r_3}={\bf 6}, \quad  {\bf r_4}={\bf 7}, \quad  {\bf r_5}={\bf 8} \ .
$$ 
The decomposition of the $SU(3)$ irreps into these has been worked out in
Ref.~\cite{Luhn:2007yr}. For the smallest irreps we have:

{\small
\begin{center}
{{\begin{tabular}{|c|}
 \hline  \\[-2mm]
~~$\bs{ SU(3)~ \supset~ \m P \m S \m L_2(7)}$  \hfill\\
   \\[-2mm]
\hline    
 \\[-2mm]
 $(10):~{\bf 3}~=~{\bf 3}\hfill$ \\ 
 $(01):~{\bf \overline 3}~=~{\bf  \overline 3}\hfill$ \\
 $(20):~{\bf 6}~=~{\bf 6}\hfill$ \\
 $(02):~{\bf \overline 6}~=~{\bf 6}\hfill$ \\
$(11):~{\bf 8}~=~{\bf 8}\hfill$ \\ 
$ (30):~{\bf 10}~=~{\bf \overline 3}+{\bf 7}\hfill$ \\
$ (21):~{\bf 15}~=~{\bf 7}+{\bf 8}\hfill$ \\
$ (40):~{\bf 15'}\:\!=~{\bf 1}+{\bf 6}+{\bf 8}\hfill$ \\
$ (05):~{\bf 21}~=~{\bf 3}+{\bf\overline 3}+{\bf 7}+{\bf 8}\hfill$ \\
$ (13):~{\bf 24}~=~{\bf\overline 3}+{\bf 6}+{\bf 7}+{\bf 8}\hfill$ \\
$ (22):~{\bf 27}~=~{\bf 6}+{\bf 6}+{\bf 7}+{\bf 8}\hfill$ \\[2mm] \hline
\end{tabular}}}\end{center}
}

Due to this decomposition, the discrete indices of most $\m{PSL}_2(7)$ irreps
${\bf r_i}$ can be simply set equal to the indices of the corresponding $SU(3)$
irreps. Since both, the ${\bf 6}$ and the $\ol{\bf 6}$ of $SU(3)$ decompose
into the ${\bf 6}$ of $\m{PSL}_2(7)$, we already see that the cubic index $\wt
A({\bf r_i})$ can only be defined modulo $N_A=14$. As for the ${\bf 7}$, we observe that
Eq.~(\ref{consistent}) requires
\beq
I({\bf 10}) ~=~\wt I(\ol{\bf 3}) \,+\, \wt I({\bf 7}) ~\mathrm{mod~}N_I \ ,
\eeq
thus fixing $\wt I({\bf 7})$ modulo~$N_I$. Having defined the values for all
$\wt I({\bf r_i})$, one can easily determine $N_I$ from the higher irreps of
$SU(3)$. For the quadratic index we obtain from the $\bf 15'$ that $N_\ell=24$.
These integers and the discrete indices of the $\m{PSL}_2(7)$ irreps are
listed in Table~\ref{tabPSL}(a) at the end of this section. 

Before proving that our assignments are consistent with Eq.~(\ref{consistent})
{\it for all} irreps ${\bs \rho}$ of $SU(3)$, we consider three examples.

\begin{itemize}
\item[(i)] ${\bs \rho} = {\bf 3}$. This is a trivial case, since the ${\bf 3}$ of
  $SU(3)$   corresponds to the ${\bf 3}$ of $\m{PSL}_2(7)$. Eq.~(\ref{consistent}) then  reads
\beqn
1 \, = \, \ell({\bf 3}) &=& \wt \ell({\bf 3}) ~\mathrm{mod~24}  \, = \,1 \
, \notag \\[1mm]
1 \, = \, A({\bf 3}) &=& \wt A({\bf 3}) ~\mathrm{mod~14}  \, = \,1 \ . \notag 
\eeqn
\item[(ii)] ${\bs \rho} = \ol{\bf 3}$. Similar to (i) we get
\beqn
1 \, = \, \ell(\ol{\bf 3}) &=& \wt \ell(\ol{\bf 3}) ~\mathrm{mod~24}  \, = \,1 \
, \notag \\[1mm]
- 1 \, = \, A(\ol{\bf 3}) &=& \wt A(\ol{\bf 3}) ~\mathrm{mod~14}  \, = \,-1 \ . \notag 
\eeqn
\item[(iii)] ${\bs \rho} = {\bf 27}$. This representation of $SU(3)$ decomposes into
  {\bf 6}+{\bf 6}+{\bf 7}+{\bf 8} of $\m{PSL}_2(7)$. Inserting the discrete
  indices of Table~\ref{tabPSL}(a) into Eq.~(\ref{consistent}) we obtain
\beqn
54 \, = \, \ell({\bf 27}) &=& 2 \,\wt \ell({\bf 6}) + \wt \ell({\bf 7}) + \wt
\ell({\bf 8})~\mathrm{mod~24}  \, = \,30~\mathrm{mod~24} \ , \notag \\[1mm]
0 \, = \, A({\bf 27}) &=& 2 \, \wt A({\bf 6}) + \wt A({\bf 7}) + \wt
A({\bf 8})~\mathrm{mod~14}  \, = \,14 ~\mathrm{mod~14} \ . \notag 
\eeqn
\end{itemize}

The first two examples serve as the basis\footnote{See Appendix~\ref{young},
  in particular the Young tableaux of Eq.~(\ref{you}) with $k=1$.} of our
proof of Eq.~(\ref{consistent}) for the group $\m{PSL}_2(7)$. As discussed in
Section~\ref{general} and Appendix~\ref{young}, the inductive step consists in
showing that Eq.~(\ref{effs}) holds true for 
${\bs \sigma}={\bf 3}, \ol{\bf 3}$:

\begin{itemize}
\item The left-hand side, i.e. the factors $f_I^{\,i}({\bs \sigma})$, can be
  calculated easily using only the information in Table~\ref{tabPSL}(a). We have
\beqn
f_I^{\,i}({\bf 3}) &=& d({\bf r_i}) \,+\, 3\, \wt I({\bf r_i}) \ ,  \label{l3}\\[1mm]
f_I^{\,i}(\ol{\bf 3}) &=& (-1)^\kappa \, d({\bf r_i}) \,+\, 3\, \wt I({\bf r_i}) \ ,\label{l3bar} 
\eeqn
with $\kappa=0$ (or 2) for the quadratic index $I=\ell$, whereas $\kappa=1$
(or 3) for the cubic index $I=A$. The explicit values for both types of
indices and all six irreps ${\bf r_i}$ of $\m{PSL}_2(7)$ are given in
the first table of Appendix \ref{f-proof}.  
\item In order to calculate the right-hand side, i.e. the factors 
\beq
\mathfrak  f_I^{\,i}({\bs \sigma}) ~=~ \wt I({\bf r_i} \otimes b_j  {\bf r_j}) \ ,\label{rsigma}
\eeq
we need to know the Kronecker products of the finite group. For $\m{PSL}_2(7)$
they can be found in Ref.~\cite{Luhn:2007yr}. Since ${\bs \sigma}$ is
constrained to be either ${\bf 3}$ or $\ol{\bf 3}$, only the following subset
of all Kronecker products is necessary. 
{\small
\begin{center}
{{\begin{tabular}{|c|}
 \hline  \\[-2mm]
~{\bf Relevant $\bs {\m P\m S\m L_2(7)}$  Kronecker Products}\hfill\\
   \\[-2mm]
\hline    
 \\[-2mm]
${\bf 3}~\otimes {\bf 3}~=~{\bf\overline 3}^{}_a~+~{\bf 6}^{}_s\hfill$ \\
${\bf 3}~\otimes {\bf\overline 3}~=~{\bf  1}~+~{\bf 8}\hfill$ \\
$\ol{\bf 3}~\otimes \ol{\bf 3}~=~{\bf3}^{}_a~+~{\bf 6}^{}_s\hfill$ \\
\\[-2mm]
$ {\bf 3}~\otimes {\bf 6}~=~{\bf\overline 3}~+~{\bf 7}~+~{\bf 8}\hfill$ \\ 
${\bf \overline 3}~\otimes {\bf 6}~=~{\bf 3}~+~{\bf 7}~+~{\bf 8}\hfill$ \\
${\bf 3}~\otimes {\bf 7}~=~{\bf 6}~+~{\bf 7}~+~{\bf 8}\hfill$ \\
${\bf \overline 3}~\otimes {\bf 7}~=~{\bf 6}~ +~{\bf 7}~ +~{\bf 8}\hfill$ \\
${\bf 3}~\otimes {\bf 8}~=~{\bf 3}~+~{\bf 6}~+~{\bf 7}~+~{\bf 8}\hfill$ \\
${\bf \overline 3}~\otimes {\bf 8}~=~{\bf\overline 3}~ +~{\bf 6}~ +~{\bf 7}~ +~{\bf 8}\hfill$ \\[1mm]
\hline
\end{tabular}}}\end{center}
}
\vskip2mm
Thus the factors $\mathfrak f_I^{\,i}({\bs \sigma})$ can be readily determined
from 
\beqn
a_i \, \mathfrak f_I^{\,i}({\bf 3}) &\!=\!& a_i \, \wt I({\bf r_i} \otimes {\bf 3}) \notag \\
&\!=\!&
a_0 \, \wt I({\bf 3}) 
\,+\, a_1  \left[\wt I(\ol{\bf 3})+ \wt I({\bf 6})\right] 
+\, a_2  \left[\wt I({\bf 1}) + \wt I({\bf 8})\right]
+\,  a_3 \left[\wt I(\ol{\bf 3}) + \wt I({\bf 7}) + \wt I({\bf 8})\right]
\notag \\
&& +\, a_4 \left[\wt I({\bf 6}) + \wt I({\bf 7}) + \wt I({\bf 8})\right]
+\,a_5 \left[\wt I({\bf 3}) + \wt I({\bf 6}) + \wt I({\bf 7})
+ \wt I({\bf 8})\right] \ ,\label{ffrak3} \\[2mm]
a_i \,\mathfrak f_I^{\,i}(\ol{\bf 3})&\!=\!& a_i\,\wt I({\bf r_i}\otimes\ol{\bf 3})\notag \\
&\!=\!&
a_0 \, \wt I(\ol{\bf 3}) 
\,+\, a_1  \left[\wt I({\bf 1})+ \wt I({\bf 8})\right]
+\, a_2  \left[\wt I({\bf 3}) + \wt I({\bf 6})\right]
+\,  a_3 \left[\wt I({\bf 3}) + \wt I({\bf 7}) + \wt I({\bf 8})\right]
\notag \\ 
&& +\, a_4 \left[\wt I({\bf 6}) + \wt I({\bf 7}) + \wt I({\bf 8})\right] 
+\,a_5 \left[\wt I(\ol{\bf 3}) + \wt I({\bf 6}) + \wt I({\bf 7})
+ \wt I({\bf 8})\right] \ ,\label{ffrak3bar}
\eeqn
for both types of indices $I=\ell,A$. Their values are calculated
and tabulated in Appendix~\ref{f-proof}.  
\end{itemize}

Having obtained the factors $f_I^{\,i}({\bs \sigma})$ and $\mathfrak
f_I^{\,i}({\bs \sigma})$ numerically, we can compare them one by one. Bearing
in mind that our calculations are only modulo $N_I$, we find that
Eq.~(\ref{effs}) is truly valid for both the quadratic as well as the cubic
index. Furthermore, the comparison also reveals that our values for $N_I$ are
the maximally allowed ones. Of course, all statements in this section would
remain true if one were to replace all $N_I$ by $N'_I=N_I/p$ where $p$ is an
integer. For instance, the cubic index could be defined modulo~$7$ instead of
modulo~$14$. This completes our proof of Eq.~(\ref{consistent}) for the group 
$\m{PSL}_2(7)$, with the discrete indices given in Table~\ref{tabPSL}(a).

\subsection{The Group $\bs{\m Z_7 \rtimes \m Z_3}$}
$\m{PSL}_2(7)$ has two maximal subgroups, one of which is the Frobenius group
$\m Z_7 \rtimes \m Z_3$, see e.g. Ref.~\cite{Luhn:2007sy}. It has the
following five irreps ${\bf r_i}$. 
$$
{\bf r_0}={\bf 1}, \quad {\bf r_1}={\bf 1'}, \quad  {\bf r_2}=\ol{\bf 1'}, \quad  {\bf r_3}={\bf 3}, \quad  {\bf r_4}=\ol{\bf 3} \ .
$$
The decomposition of $SU(3)$ irreps into these can be easily obtained from the
embedding sequence $SU(3) \supset \m{PSL}_2(7) \supset \m Z_7 \rtimes \m Z_3$
\cite{Luhn:2007yr}, yielding the result:

{\small
\begin{center}
{{\begin{tabular}{|c|}
 \hline  \\[-2mm]
~~$\bs{ SU(3)~ \supset~ \m Z_7 \rtimes  \m Z_3}$  \hfill\\
   \\[-2mm]
\hline    
 \\[-2mm]
 $(10):~{\bf 3}~=~{\bf 3}\hfill$ \\ 
 $(01):~{\bf \overline 3}~=~{\bf  \overline 3}\hfill$ \\
 $(20):~{\bf 6}~=~{\bf 3} + \ol{\bf 3}\hfill$ \\
 $(02):~{\bf \overline 6}~=~{\bf 3} + \ol{\bf 3}\hfill$ \\
$(11):~{\bf 8}~=~{\bf 1'}+\ol{\bf 1'}+{\bf 3}+\ol{\bf 3}\hfill$ \\ 
$ (30):~{\bf 10}~=~{\bf 1}+{\bf 3}+2\cdot\ol{\bf 3}\hfill$ \\
$ (21):~{\bf 15}~=~{\bf 1}+{\bf 1'}+\ol{\bf 1'}+2\cdot({\bf 3}+\ol{\bf 3})\hfill$ \\
$ (40):~{\bf 15'}\;\!=~{\bf 1}+{\bf 1'}+\ol{\bf 1'}+2\cdot({\bf 3}+\ol{\bf 3})\hfill$ \\
$ (05):~{\bf 21}~=~{\bf 1}+{\bf 1'}+\ol{\bf 1'}+3\cdot({\bf 3}+\ol{\bf 3})\hfill$ \\
$ (13):~{\bf 24}~=~{\bf 1}+{\bf 1'}+\ol{\bf 1'}+3\cdot{\bf 3}+4\cdot\ol{\bf 3}\hfill$ \\
$ (22):~{\bf 27}~=~{\bf 1}+{\bf 1'}+\ol{\bf 1'}+4\cdot({\bf 3}+\ol{\bf 3})\hfill$ \\[2mm] \hline
\end{tabular}}}\end{center}
}
\noindent Notice that ${\bf 1'}$ and $\ol{\bf 1'}$ always come in pairs in the
decomposition of the smallest $SU(3)$ irreps. It is easy to prove this
peculiarity for arbitrary irreps of $SU(3)$ by induction. Assume that ${\bs
  \rho}$ decomposes as 
\beq
{\bs \rho}~=~a_0\,{\bf 1} ~+~a_1\,({\bf 1'} \:+\:\ol{\bf 1'}) ~+~a_3\,{\bf 3}
~+~a_4\,\ol{\bf 3} \ .\label{together1}
\eeq
Using Eq.~(\ref{decKro}) and the $\m Z_7\rtimes \m Z_3$ Kronecker products
\cite{Luhn:2007yr} 
{\small
\begin{center}
{{\begin{tabular}{|c|}
 \hline  \\[-2mm]
{\bf $\bs {\m Z_7 \rtimes \m Z_3}$  Kronecker Products}\hfill\\
   \\[-2mm]
\hline    
 \\[-2mm]
${\bf 1'}\;\!\otimes \,{\bf 1'}\,=~\ol{\bf 1'}\hfill$ \\
${\bf 1'}\;\!\otimes \,\ol{\bf 1'}\,=~{\bf 1}\hfill$ \\
${\bf 3}~\otimes \,{\bf 1'}\,=~{\bf 3}\hfill$ \\
${\bf 3}~\otimes \,\ol{\bf 1'}\,=~{\bf 3}\hfill$ \\
${\bf 3}~\otimes \;{\bf 3}~=~({\bf  3}~+~\ol{\bf 3})_s ~+~\ol{\bf 3}_a\hfill$ \\
${\bf 3}~\otimes \;\ol{\bf 3}~=~{\bf  1}~+~ {\bf  1'}~+~\ol{\bf  1'}~+~  {\bf 3}
~+~\ol{\bf 3}\hfill$  \\[2mm]
\hline
\end{tabular}}}\end{center}
}
\noindent we obtain the decomposition of the representations ${\bs \rho}
\otimes {\bf 3}$ and  ${\bs \rho} \otimes \ol{\bf 3}$.
\beqn
{\bs \rho} \otimes {\bf 3} &=&a_4\,{\bf 1} ~+~ a_4\,({\bf 1'}\:+\:\ol{\bf 1'}) ~+~
(a_0+2a_1+a_3+a_4)\,{\bf 3} ~+~ (2a_3+a_4)\,\ol{\bf 3} \ ,\notag \\[1mm]
{\bs \rho} \otimes \ol{\bf 3} &=&a_3\,{\bf 1} ~+~ a_3\,({\bf 1'}\:+\:\ol{\bf 1'}) ~+~
(a_3+2a_4)\,{\bf 3} ~+~ (a_0+2a_1+a_3+a_4)\,\ol{\bf 3} \ . \notag
\eeqn
Of course, these are the decompositions of {\it reducible} $SU(3)$
representations, i.e. of sums of $SU(3)$ irreps. As argued in
Appendix~\ref{young}, such a sum contains {\it only one new} $SU(3)$
irrep. Assuming that the other known irreps decompose with the  ${\bf 1'}$ and
$\ol{\bf 1'}$ appearing in pairs, this is true also for the new $SU(3)$ irrep
and therefore for all.

Since Eq.~(\ref{together1}) holds for any irrep ${\bs \rho}$, the discrete
indices cannot be defined uniquely. We take this fact into account by
introducing the parameters $x$ and $y$. For physical applications of the
discrete indices, it might be convenient to choose specific values, see
Section~\ref{DACsect}. At this point, however, we want to stay as general as
possible, thus leaving $x$ and $y$ undetermined. It should also be stressed
that there is nothing wrong with having non-integer
values. Table~\ref{tabPSL}(b) shows the discrete indices of $\m Z_7 \rtimes \m
Z_3 \subset SU(3)$. The values of $N_I$ for both types of indices are determined
by the decomposition of the ${\bf 6}$.

This assignment trivially satisfies Eq.~(\ref{consistent}) for ${\bs \rho}={\bf
  3},\ol{\bf 3}$. In order to prove it for all other $SU(3)$ irreps, we need
to compare $f_I^{\,i}({\bs \sigma})$ and $\mathfrak f_I^{\,i}({\bs \sigma})$
for ${\bs \sigma}={\bf 3},\ol{\bf 3}$.  The former, i.e. $f_I^{\,i}({\bs
  \sigma})$, is calculated from  Eqs.~(\ref{l3}) and~(\ref{l3bar}) using
Table~\ref{tabPSL}(b). $\mathfrak f_I^{\,i}({\bs \sigma})$ on the other hand
is determined from Eq.~(\ref{rsigma}) with the Kronecker products and the
discrete indices of $\m Z_7 \rtimes \m Z_3$.  
Their explicit values for both types of indices are listed in
Appendix~\ref{f-proof}. Note that we only need to compare the {sum}
$f_I^{\,1+2}({\bs \sigma})=f_I^{\,1}({\bs \sigma})+f_I^{\,2}({\bs \sigma})$
with the {sum} $\mathfrak f_I^{\,1+2}({\bs \sigma})=\mathfrak f_I^{\,1}({\bs
  \sigma})+ \mathfrak  f_I^{\,2}({\bs \sigma})$ because ${\bf 1'}$ and
$\ol{\bf 1'}$ come in pairs in the decomposition of any $SU(3)$ irrep ${\bs
  \rho}$.  This comparison shows that our definition of the discrete indices
for the group $\m Z_7 \rtimes \m Z_3$, given in Table~\ref{tabPSL}(b),
satisfies Eq.~(\ref{consistent}) and is therefore consistent. 

\vspace{3mm}

\begin{table}[h]
\subtable[Discrete indices of $\m{PSL}_2(7) \subset SU(3)$.]
{
~~~\begin{tabular}{|c|r|r|}
\hline ~ && \\[-3mm]
{\begin{tabular}{c} $\m{PSL}_2(7)$  \\[1mm] irreps \end{tabular}} 
& $\!\!\begin{array}{c} \wt \ell({\bf r}) \\[1mm] (N_\ell\,=\,24) \end{array}\!\!$ 
&  $\!\!\begin{array}{c} \wt A({\bf r})\\[1mm](N_A\,=\,14) \end{array}\!\!$ 
\\[-3mm]~&& \\\hline 
~&&\\[-3mm]
{\bf 1}       &  0~~~~~~   &      0~~~~~~ \\
{\bf 3}       &  1~~~~~~   &      1~~~~~~ \\
$\ol{\bf 3}$  &  1~~~~~~   &   $-1$~~~~~~ \\
{\bf 6}       &  5~~~~~~   &      7~~~~~~ \\
{\bf 7}  &  14~~~~~~   &   0~~~~~~  \\
{\bf 8}       &  6~~~~~~   &      0~~~~~~ \\[1mm]\hline
\end{tabular} 
}
\subtable[Discrete indices of $\m Z_7 \rtimes \m Z_3 \subset SU(3)$.]{
~~\begin{tabular}{|c|r|r|}
\hline ~ && \\[-3mm]
{\begin{tabular}{c} $\m Z_7\rtimes\m Z_3$ \\[1mm] irreps \end{tabular}} 
& $\!\!\begin{array}{c} \wt \ell({\bf r}) \\[1mm] (N_\ell\,=\,3) \end{array}\!\!$ 
&  $\!\!\begin{array}{c} \wt A({\bf r})\\[1mm](N_A\,=\,7) \end{array}\!\!$ 
\\[-3mm]~&& \\\hline 
~&&\\[-3mm]
{\bf 1}$\:$       &      0~~~~~~   &          0~~~~~~ \\
${\bf 1'}$    &    $x$~~~~~~   &        $y$~~~~~~ \\
$\ol{\bf 1'}$ &  $1-x$~~~~~~   &       $-y$~~~~~~ \\
{\bf 3}$\;$       &      1~~~~~~   &          1~~~~~~ \\
$\ol{\bf 3}\;$  &      1~~~~~~   &       $-1$~~~~~~ \\[1mm]\hline
\end{tabular}
}  
\caption{\label{tabPSL}The definition of the discrete indices of the finite
  groups $\m{PSL}_2(7)$ and $\m Z_7  \rtimes \m Z_3$ originating in the
  continuous group $SU(3)$. $x$ and $y$ can take arbitrary values.}  
\end{table}

\section{\label{othersect}Indices of $\bs{\Delta(27)}$, $\bs{\m S_4}$, $\bs{\m A_4}$, $\bs{\m  D_5}$, and $\bs{\m S_3}$}
\cleqn
In this section we discuss the discrete indices for other finite subgroups of
$SU(3)$. To be self-contained, we also list the embedding of their irreps into
those of $SU(3)$, as well as their Kronecker products. 
However, we refrain from showing explicitly that our definitions of discrete
indices are consistent with Eq.~(\ref{consistent}). This can be done
analogously to the previous sections. Our results for the discrete indices are
presented in Table~\ref{tabSUM} at the end of this section.

\subsection{The Group $\bs{\Delta(27)}$}
This group, see e.g. Ref.~\cite{Luhn:2007uq}, has nine one-dimensional irreps,
${\bf 1_{r,s}}$ with $r,s=0,1,2$, as well as two three-dimensional ones, ${\bf
  3}$ and $\ol{\bf 3}$. For the one-dimensional irreps we also write
\vspace{1mm}$$
\begin{array}{lllll}
{\bf 1_0}={\bf 1_{0,0}}\,,
& ~~ {\bf 1_1}={\bf 1_{0,1}}\,,&~~{\bf 1_3}={\bf 1_{1,0}}\,,&~~{\bf
  1_5}={\bf 1_{1,1}}\,,&~~{\bf 1_7}={\bf 1_{1,2}}\,, \\[4mm]
& ~~ {\bf 1_2}={\bf \ol 1_1}={\bf 1_{0,2}}\,,&~~{\bf 1_4}={\bf \ol 1_3}={\bf
  1_{2,0}}\,,&~~{\bf 1_6}={\bf \ol 1_5}={\bf 1_{2,2}}\,,&~~{\bf 1_8}={\bf \ol
  1_7}={\bf 1_{2,1}}\,. 
\end{array}
$$

\vspace{2mm}

\noindent With this notation, the Kronecker products and the decomposition of
the smallest $SU(3)$ irreps are given as:
{\small
\begin{center}
\begin{tabular}{ccc}
\begin{tabular}{|c|}
 \hline  \\[-2mm]
{\bf $\bs {\Delta(27)}$  Kronecker Products}\hfill\\
   \\[-2mm]
\hline    
 \\[-2mm]
${\bf 1_{r,s}}\;\!\otimes \,{\bf 1_{r',s'}}\,=~{\bf 1_{r+r',s+s'}}\hfill$ \\
${\bf 3}~\otimes \,{\bf 1_j}\,=~{\bf 3}\hfill$ \\
$\ol{\bf 3}~\otimes \,{\bf 1_j}\,=~\ol{\bf 3}\hfill$ \\
${\bf 3}~\otimes \;{\bf 3}~=~3\cdot \ol{\bf  3}\hfill$ \\
$\ol{\bf 3}~\otimes \;\ol{\bf 3}~=~3\cdot {\bf  3}\hfill$  \\
${\bf 3}~\otimes \;\ol{\bf 3}~=~{\bf 1_0} + \sum_{j=1}^8 {\bf  1_j}\hfill$  \\[2mm]
\hline
\end{tabular}
&&
\begin{tabular}{|c|}
 \hline  \\[-2mm]
~~$\bs{ SU(3)~ \supset~ \Delta(27)}$  \hfill\\
   \\[-2mm]
\hline    
 \\[-2mm]
 $(10):~{\bf 3}~=~{\bf 3}\hfill$ \\ 
 $(01):~{\bf \overline 3}~=~{\bf  \overline 3}\hfill$ \\
 $(20):~{\bf 6}~=~2\cdot \ol{\bf 3}\hfill$ \\
 $(02):~{\bf \overline 6}~=~2\cdot{\bf 3} \hfill$ \\
$(11):~{\bf 8}~=~\sum_{j=1}^8 {\bf 1_j}\hfill$ \\ 
$ (30):~{\bf 10}~=~2\cdot {\bf 1_0} + ~\sum_{j=1}^8 {\bf 1_j}\hfill$ \\
$ (21):~{\bf 15}~=~5\cdot {\bf 3} \hfill$ \\
$ (40):~{\bf 15'}\;\!=~5\cdot {\bf 3}   \hfill$ \\
$ (05):~{\bf 21}~=~ 7\cdot {\bf 3}  \hfill$ \\
$ (13):~{\bf 24}~=~8\cdot {\bf 3}   \hfill$ \\
$ (22):~{\bf 27}~=~ 3\cdot ({\bf 1_0} + \sum_{j=1}^8 {\bf 1_j})\hfill$ \\[2mm] \hline
\end{tabular}
\end{tabular}\end{center}
}

\vspace{2mm}

\noindent On the right-hand side of the Kronecker product for the
one-dimensional irreps, the sums $r+r'$  and $s+s'$ are modulo~3.
Similar to the $\m Z_7 \rtimes \m Z_3$ case, one can easily show that the
one-dimensional irreps ${\bf 1_j}$ with $j=1,...,8$ occur always collectively
in the decomposition of $SU(3)$ irreps. The resulting ambiguity in the
definition of the corresponding discrete indices is expressed by introducing
the parameters $x_k$ and $y_k$ with $k=1,...,7$ in
Table~\ref{tabSUM}(a). The decomposition of the ${\bf 6}$ fixes the
values of $N_I$.


\subsection{The Group $\bs{\m S_4}$}
Besides $\m Z_7 \rtimes \m Z_3$, this group is the second maximal subgroup of
$\m{PSL}_2(7)$. It has five irreps.
\vspace{2mm}$$
{\bf r_0}={\bf 1}, \quad {\bf r_1}={\bf 1'}, \quad {\bf r_2}={\bf 2}, \quad {\bf r_3}={\bf 3_1}, \quad
{\bf r_4}={\bf 3_2} \ .
$$

\vspace{2mm}

\noindent The Kronecker products of $\m S_4$ and its embedding into $SU(3)$ are~\cite{Luhn:2007yr,Hagedorn:2006ug}:
{\small
\begin{center}
\begin{tabular}{ccc}
\begin{tabular}{|c|}
 \hline  \\[-2mm]
{\bf $\bs {\m S_4}$  Kronecker Products}\hfill\\
   \\[-2mm]
\hline    
 \\[-2mm]
${\bf 1'}\:\otimes \,{\bf 1'}\,=~{\bf 1}\hfill$ \\
${\bf 2}\;\,\otimes \,{\bf 1'}\,=~{\bf 2}\hfill$ \\
${\bf 3_1}\otimes \,{\bf 1'}\,=~{\bf 3_2}\hfill$ \\
${\bf 3_2}\otimes \,{\bf 1'}\,=~{\bf 3_1}\hfill$ \\
${\bf 2}\;\,\otimes \,{\bf 2}\;\,=~({\bf 1} + {\bf 2})_s + ({\bf 1'})_a \hfill$ \\
${\bf 2}\;\,\otimes \,{\bf 3_i}\,=~{\bf 3_1} + {\bf 3_2} \hfill$ \\
${\bf 3_i}\,\,\!\otimes \,{\bf 3_i}\,=~({\bf 1}+{\bf 2}+{\bf 3_1})_s+({\bf 3_2})_a \hfill$ \\
${\bf 3_1}\otimes \,{\bf 3_2}=~{\bf  1'}+{\bf  2}+{\bf  3_1}+{\bf  3_2}\hfill$  \\[2mm]
\hline
\end{tabular}
&&
\begin{tabular}{|c|}
 \hline  \\[-2mm]
~~$\bs{ SU(3)~ \supset~ \m S_4}$  \hfill\\
   \\[-2mm]
\hline    
 \\[-2mm]
 $(10):~{\bf 3}~=~{\bf 3_2}\hfill$ \\ 
 $(01):~{\bf \overline 3}~=~{\bf  3_2}\hfill$ \\
 $(20):~{\bf 6}~=~{\bf 1}+{\bf 2} + {\bf 3_1}\hfill$ \\
 $(02):~{\bf \overline 6}~=~ {\bf 1}+{\bf 2} + {\bf 3_1}\hfill$ \\
$(11):~{\bf 8}~=~ {\bf 2}+{\bf 3_1} + {\bf 3_2}  \hfill$ \\ 
$ (30):~{\bf 10}~=~  {\bf 1'}+{\bf 3_1} + 2\!\cdot\! {\bf 3_2} \hfill$ \\
$ (21):~{\bf 15}~=~  {\bf 1'}+{\bf 2}+2\!\cdot\!({\bf 3_1}+{\bf 3_2})\hfill$ \\
$ (40):~{\bf 15'}\;\!=~2\!\cdot\!({\bf 1}+{\bf 2}+{\bf 3_1})+{\bf 3_2}\hfill$ \\
$ (05):~{\bf 21}~=~ {\bf 1'}+{\bf 2}+2\!\cdot\!{\bf 3_1}+4\!\cdot\!{\bf 3_2} 
\hfill$ \\
$ (13):~{\bf 24}~=~{\bf 1}+{\bf 1'}+2\!\cdot\!{\bf 2}+3\!\cdot\!({\bf 3_1}
+{\bf 3_2}) \hfill$ \\
$ (22):~{\bf 27}~=~ 2\!\cdot\! {\bf 1}+{\bf 1'}+3\!\cdot\!{\bf 2}+4\!\cdot\!
{\bf 3_1} + 2\!\cdot\! {\bf 3_2}   \hfill$ \\[2mm] \hline
\end{tabular}
\end{tabular}\end{center}
}

\vspace{2mm}

\noindent Notice that the occurrence of both ${\bf 1'}$ and ${\bf 2}$ is
always accompanied by the irrep ${\bf 3_1}$ in the decomposition of the smallest
$SU(3)$ irreps. Again, it is easy to prove this for all irreps of $SU(3)$ by
induction. Assume that ${\bs \rho}$ decomposes as 
\beq
{\bs \rho}~=~a_0\,{\bf 1} ~+~a_1\, ({\bf 1'}+{\bf 3_1})  ~+~ a_2 \, (\,{\bf 2}+{\bf 3_1})  ~+~ a_4\,{\bf 3_2} \ .\label{together2}
\eeq
Since ${\bf 3}$ and $\ol{\bf 3}$ both correspond to the same $\m S_4$ irrep
${\bf 3_2}$, the two $SU(3)$ representations ${\bs \rho} \otimes {\bf 3}$ and ${\bs \rho}
\otimes  \ol{\bf 3}$ have the same decomposition. It is obtained from the
Kronecker products, yielding
$$
 a_4\,{\bf 1} ~+~(a_1+a_2) \, ({\bf 1'}+{\bf 3_1})  ~+~ (a_1+a_2+a_4) \, (\,{\bf 2}+{\bf 3_1})  ~+~ (a_0+a_1+2a_2+a_4)\,{\bf 3_2}        \ ,
$$
which is of the same structure as Eq.~(\ref{together2}). Due to this general
property of the embedding of $\m S_4$ into $SU(3)$, the discrete indices are
not defined uniquely. The values for $\wt I({\bf 2+3_1})$ and $\wt I({\bf
  1'+3_1})$ are given by the ${\bf 6}$ and the ${\bf 10}$ of $SU(3)$,
respectively. The ${\bf 15'}$ then determines $N_\ell$ to be 24 for the
quadratic index, while $N_A=2$  for the cubic index because both the ${\bf 3}$
and  the  $\ol{\bf 3}$ decompose as a ${\bf 3_2}$ of $\m S_4$. The results are
shown in Table~\ref{tabSUM}(b), with the ambiguity in the definitions
parameterized by $x$ and $y$.


\subsection{The Group $\bs{\m A_4}$}
This group is the most popular group in flavor model building. It is a
subgroup of $\m S_4$ and has four irreps.
\vspace{2mm}$$
{\bf r_0}={\bf 1}, \quad {\bf r_1}={\bf 1'}, \quad {\bf r_2}=\ol{\bf 1'}, \quad {\bf r_3}={\bf 3} \ .
$$

\vspace{2mm}

\noindent The Kronecker products of $\m A_4$ are listed throughout the
literature. The decomposition of the smallest $SU(3)$ irreps can be worked out
easily from $\m A_4$'s embedding in $\m S_4$~\cite{Luhn:2007yr}.
{\small
\begin{center}
\begin{tabular}{ccc}
\begin{tabular}{|c|}
 \hline  \\[-2mm]
{\bf $\bs {\m A_4}$  Kronecker Products}\hfill\\
   \\[-2mm]
\hline    
 \\[-2mm]
${\bf 1'}\otimes \,{\bf 1'}=~\ol{\bf 1'}\hfill$ \\
${\bf 1'}\otimes \,\ol{\bf 1'}=~{\bf 1}\hfill$ \\
${\bf 3}\:\otimes \,{\bf 1'}=~{\bf 3}\hfill$ \\
${\bf 3}\:\otimes\,{\bf 3}\:=~{\bf 1}+{\bf 1'}+\ol{\bf 1'}+2\cdot{\bf 3}\hfill$  \\[2mm]
\hline
\end{tabular}
&&
\begin{tabular}{|c|}
 \hline  \\[-2mm]
~~$\bs{ SU(3)~ \supset~ \m A_4}$  \hfill\\
   \\[-2mm]
\hline    
 \\[-2mm]
 $(10):~{\bf 3}~=~{\bf 3}\hfill$ \\ 
 $(01):~{\bf \overline 3}~=~{\bf  3}\hfill$ \\
 $(20):~{\bf 6}~=~{\bf 1}+{\bf 1'}+\ol{\bf 1'} + {\bf 3}\hfill$ \\
 $(02):~{\bf \overline 6}~=~ {\bf 1}+{\bf 1'}+\ol{\bf 1'} + {\bf 3}   \hfill$ \\
$(11):~{\bf 8}~=~ {\bf 1'}+\ol{\bf 1'} + 2\cdot {\bf 3} \hfill$ \\ 
$ (30):~{\bf 10}~=~ {\bf 1} + 3\cdot {\bf 3} \hfill$ \\
$ (21):~{\bf 15}~=~  {\bf 1}+{\bf 1'}+\ol{\bf 1'} + 4\cdot{\bf 3} \hfill$ \\
$ (40):~{\bf 15'}\;\!=~ 2\cdot({\bf 1}+{\bf 1'}+\ol{\bf 1'}) + 3\cdot{\bf 3} \hfill$ \\
$ (05):~{\bf 21}~=~ {\bf 1}+{\bf 1'}+\ol{\bf 1'} + 6\cdot{\bf 3}  \hfill$ \\
$ (13):~{\bf 24}~=~ 2\cdot({\bf 1}+{\bf 1'}+\ol{\bf 1'}) + 6\cdot{\bf 3} \hfill$ \\
$ (22):~{\bf 27}~=~ 3\cdot({\bf 1}+{\bf 1'}+\ol{\bf 1'}) + 6\cdot{\bf 3}   \hfill$ \\[2mm] \hline
\end{tabular}
\end{tabular}\end{center}
}

\noindent In this case, the irreps ${\bf 1'}$ and $\ol{\bf 1'}$ always come in
pairs in the decomposition of the $SU(3)$ irreps. The discrete indices, listed
in Table~\ref{tabSUM}(c), are again not uniquely determined. As before
$N_A=2$, while the ${\bf 10}$ fixes $N_\ell$ to be 12. 


\subsection{The Group $\bs{\m D_5}$}
The dihedral group $\m D_5$ has also been used as a family group. 
Its four irreps are
$$
{\bf r_0}={\bf 1}, \quad {\bf r_1}={\bf 1'}, \quad {\bf r_2}={\bf 2_1}, \quad {\bf r_3}={\bf 2_2} \ .
$$
The Kronecker products of $\m D_5$ as well as its embedding in $SU(3)$ can be
found in Refs.~\cite{Hagedorn:2006ir}.
{\small
\begin{center}
\begin{tabular}{ccc}
\begin{tabular}{|c|}
 \hline  \\[-2mm]
{\bf $\bs {\m D_5}$  Kronecker Products}\hfill\\
   \\[-2mm]
\hline    
 \\[-2mm]
${\bf 1'}\:\otimes {\bf 1'}\,=~{\bf 1}\hfill$ \\
${\bf 1'}\:\otimes{\bf 2_i}\,=~{\bf 2_i}\hfill$ \\
${\bf 2_1}\otimes {\bf 2_1}=~{\bf 1} +{\bf 1'} +{\bf 2_2} \hfill$ \\
${\bf 2_2}\otimes {\bf 2_2}=~{\bf 1} +{\bf 1'} +{\bf 2_1} \hfill$ \\
${\bf 2_1}\otimes {\bf 2_2}=~{\bf 2_1}+{\bf 2_2} \hfill$ \\[2mm]
\hline
\end{tabular}
&&
\begin{tabular}{|c|}
 \hline  \\[-2mm]
~~$\bs{ SU(3)~ \supset~ \m D_5}$  \hfill\\
   \\[-2mm]
\hline    
 \\[-2mm]
 $(10):~{\bf 3}~=~{\bf 1'}+{\bf 2_1}\hfill$ \\ 
 $(01):~{\bf \overline 3}~=~ {\bf 1'}+{\bf 2_1}  \hfill$ \\
 $(20):~{\bf 6}~=~  2\cdot{\bf 1} + {\bf 2_1}+{\bf 2_2}  \hfill$ \\
 $(02):~{\bf \overline 6}~=~2\cdot{\bf 1} + {\bf 2_1}+{\bf 2_2} \hfill$ \\
$(11):~{\bf 8}~=~ {\bf 1} +{\bf 1'} + 2\cdot{\bf 2_1}+{\bf 2_2}  \hfill$ \\ 
$ (30):~{\bf 10}~=~2\cdot( {\bf 1'} + {\bf 2_1}+{\bf 2_2})   \hfill$ \\
$ (21):~{\bf 15}~=~{\bf 1} + 2\cdot{\bf 1'} +  3\cdot({\bf 2_1}+{\bf 2_2})   \hfill$ \\
$ (40):~{\bf 15'}\;\!=~ 3\cdot( {\bf 1} + {\bf 2_1}+{\bf 2_2})    \hfill$ \\
$ (05):~{\bf 21}~=~ {\bf 1}+4\cdot( {\bf 1'} + {\bf 2_1}+{\bf 2_2})   \hfill$ \\
$ (13):~{\bf 24}~=~  2\cdot({\bf 1} +{\bf 1'})+5\cdot({\bf 2_1}+{\bf 2_2})   \hfill$ \\
$ (22):~{\bf 27}~=~4\cdot {\bf 1} + {\bf 1'} + 5\cdot {\bf 2_1} + 6\cdot {\bf 2_2}    \hfill$ \\[2mm] \hline
\end{tabular}
\end{tabular}\end{center}
}

\noindent Another alternative embedding of $\m D_5$ in $SU(3)$ is obtained by
exchanging the representations ${\bf 2_1} \leftrightarrow {\bf 2_2}$. However,
we only spell out the results for the choice shown above. First notice that
the smallest irreps ${\bs \rho}$ of $SU(3)$ all decompose as
$$
{\bs \rho}~=~a_0\,{\bf 1} ~+~a_1\,{\bf 1'} ~+~a_2\,{\bf 2_1} ~+~a_3\,{\bf 2_2} \ ,
$$
with 
$a_1+a_2+a_3$ being even. This can be verified for higher irreps by examining
the decomposition of ${\bs \rho}\otimes{\bf 3}$ and ${\bs \rho}\otimes\ol{\bf 3}$, for
both of which we find
$$
(a_1+a_2)\,{\bf 1} ~+~\underbrace{(a_0+a_2)}_{a_1'}\,{\bf 1'} ~+~\underbrace{(a_0+a_1+a_2+a_3)}_{a_2'}\,{\bf 2_1} ~+~\underbrace{(a_2+2 a_3)}_{a_3'}\,{\bf 2_2} \ .
$$
Obviously, the sum $a_1'+a_2'+a_3'$ is even again. This fact has to be taken
into account in the inductive proof of Eq.~(\ref{consistent}) where
$f_I^{\,i}({\bs \sigma})$ and $\mathfrak f_I^{\,i}({\bs \sigma})$ are
calculated and compared in a table similar to those shown in
Appendix~\ref{f-proof}. 

The assignment of the discrete indices is somewhat more involved because the
${\bf 3}$ of $SU(3)$ decomposes into two irreps of $\m D_5$. Writing 
$$
\wt I({\bf 1'}) = \alpha\ , \qquad \wt I({\bf 2_1}) = \beta\ , \qquad \wt I({\bf 2_2}) = \gamma\ ,
$$
we get
\vspace{-2mm}
\beqn
I({\bf 3})~=~ I(\ol{\bf 3})&=& \alpha ~+~\beta \mathrm{~~mod~~}N_I\ , \notag  \\
I({\bf 6}) &=& \beta ~+~\gamma\mathrm{~~mod~~}N_I\ , \notag  \\
I({\bf 15'}) &=& 3\,(\beta ~+~\gamma)\mathrm{~~mod~~}N_I\ , \notag  \\
I({\bf 27}) &=& \alpha ~+~5\,\beta ~+~6\,\gamma \mathrm{~~mod~~}N_I\ . \notag  
\eeqn
For the quadratic index, $N_\ell$ is determined by comparing the ${\bf 15'}$
with three copies of the~${\bf 6}$, yielding $N_\ell=20$. For the cubic index
we have $N_A=2$. The value of $\alpha$ can be calculated from
$$
I({\bf 3})~+~I({\bf 27})~-~6\,I({\bf 6})~=~ 2 \, \alpha \mathrm{~~mod~~}N_I\ .
$$
Due to the mod~$N_I$ we get two discrete solutions, parameterized by $\xi=0,1$
and $\zeta=0,1$, respectively. The values for $\beta$ and $\gamma$ are then
easily determined from the ${\bf 3}$ and the ${\bf 6}$. Table~\ref{tabSUM}(d)
lists the results. Notice that these indices are {\it not} integer. 


\subsection{The Group $\bs{\m S_3}$}
Finally, the last group we consider is $\m S_3$ with three irreps
$$
{\bf r_0}={\bf 1}, \quad {\bf r_1}={\bf 1'}, \quad {\bf r_2}={\bf 2} \ .
$$
The Kronecker products of $\m S_3$ and its embedding in $SU(3)$ is given below.
\vspace{1mm}

{\small
\begin{center}
\begin{tabular}{ccc}
\begin{tabular}{|c|}
 \hline  \\[-2mm]
{\bf $\bs {\m S_3}$  Kronecker Products}\hfill\\
   \\[-2mm]
\hline    
 \\[-2mm]
${\bf 1'}\otimes {\bf 1'}\,=~{\bf 1}\hfill$ \\
${\bf 1'}\otimes\:\!{\bf 2}~=~{\bf 2}\hfill$ \\
${\bf 2}\:\otimes\:\! {\bf 2}~=~{\bf 1} +{\bf 1'} +{\bf 2} \hfill$ \\[2mm]
\hline
\end{tabular}
&&
\begin{tabular}{|c|}
 \hline  \\[-2mm]
~~$\bs{ SU(3)~ \supset~ \m S_3}$  \hfill\\
   \\[-2mm]
\hline    
 \\[-2mm]
 $(10):~{\bf 3}~=~{\bf 1'}+{\bf 2}\hfill$ \\ 
 $(01):~{\bf \overline 3}~=~ {\bf 1'}+{\bf 2}  \hfill$ \\
 $(20):~{\bf 6}~=~ 2\cdot({\bf 1} + {\bf 2})  \hfill$ \\
 $(02):~{\bf \overline 6}~=~2\cdot({\bf 1} + {\bf 2}) \hfill$ \\
$(11):~{\bf 8}~=~ {\bf 1} + {\bf 1'} +  3\cdot{\bf 2}    \hfill$ \\ 
$ (30):~{\bf 10}~=~ {\bf 1} + 3\cdot({\bf 1'} + {\bf 2})  \hfill$ \\
$ (21):~{\bf 15}~=~  2\cdot{\bf 1} + 3\cdot{\bf 1'} + 5\cdot{\bf 2}  \hfill$ \\
$ (40):~{\bf 15'}\;\!=~   4\cdot{\bf 1} + {\bf 1'} + 5\cdot{\bf 2}   \hfill$ \\
$ (05):~{\bf 21}~=~   2\cdot{\bf 1} + 5\cdot{\bf 1'} + 7\cdot{\bf 2}  \hfill$ \\
$ (13):~{\bf 24}~=~  4\cdot({\bf 1} + {\bf 1'}) + 8\cdot{\bf 2}    \hfill$ \\
$ (22):~{\bf 27}~=~   6\cdot{\bf 1} + 3\cdot{\bf 1'} + 9\cdot{\bf 2}  \hfill$ \\[2mm] \hline
\end{tabular}
\end{tabular}\end{center}
}

\noindent Analogous to the group $\m D_5$, the $SU(3)$ irreps decompose into
irreps of $\m S_3$ such that the sum of the multiplicities of ${\bf 1'}$ and
${\bf 2}$ is even.

For the quadratic index, $N_\ell$ is obtained by comparing the ${\bf 10}$ with
three copies of the~${\bf 3}$, yielding $N_\ell=12$. For the cubic index we have
$N_A=2$. The ${\bf 6}$ determines $\wt I({\bf 2})$, again with two discrete
solutions parameterized by $\xi$ and $\zeta$. Then, $\wt I({\bf 1'})$ can be
calculated from the~${\bf 3}$. The resulting discrete indices of $\m S_3$ are
given in Table~\ref{tabSUM}(e).



\vspace{4mm}

\begin{table}[h]
\subfigure[\label{tabDeltak}Discrete indices of $\Delta(27) \subset SU(3)$; $k=1,...,7$.]
{
~\begin{tabular}{|c|r|r|}
\hline ~ && \\[-3mm]
{\begin{tabular}{c} $\Delta(27)$ \\[1mm] \,irreps\, \end{tabular}} 
& $\!\begin{array}{c} \wt \ell({\bf r}) \\[1mm] ~(N_\ell\,=\,3)~ \end{array}\!~$ 
&  $\!\begin{array}{c} \wt A({\bf r})\\[1mm]~(N_A\,=\,9)~ \end{array}\!~$ 
\\[-3mm]~&& \\\hline 
~&&\\[-3mm]
${\bf 1_0}$       &      0~~~\,   &          0~~~ \\
${\bf 1_k}$       &  $x_k$~~\,   &      $y_k$~~ \\
${\bf 1_8}$       & $\,-\sum_{k=1}^7 x_k$~~\,   & $-\sum_{k=1}^7 y_k$~~ \\
${\bf 3}$       &      1~~~\,   &          1~~~ \\
$\ol{\bf 3}$  &      1~~~\,   &       $-1$~~~ \\[1mm]\hline
\end{tabular}
}\quad
\subfigure[\label{tabS4k}Discrete indices of $\m S_4  \subset SU(3)$.]
{
\begin{tabular}{|c|r|r|}
\hline ~ && \\[-3mm]
{\begin{tabular}{c} $\m S_4$ \\[1mm] \,irreps\, \end{tabular}} 
& $\!\begin{array}{c} \wt \ell({\bf r}) \\[1mm] \,(N_\ell\,=\,24) \end{array}\!~$ 
&  $\!\begin{array}{c} \wt A({\bf r})\\[1mm](N_A\,=\,2) \end{array}\!~$ 
\\[-3mm]~&& \\\hline 
~&&\\[-3mm]
${\bf 1}~$       &      0~~~~~~   &           0~~~~~~ \\
${\bf 1'}$      &  $13-x$~~~~~~   &        $1-y$~~~~~~ \\
${\bf 2}~$       &  $5-x$~~~~~~   &        $1-y$~~~~~~ \\
${\bf 3_1}$     &     $x$~~~~~~   &          $y$~~~~~~ \\
${\bf 3_2}$     &       1~~~~~~   &            1~~~~~~ \\[1mm]\hline
\end{tabular}
}\\
\subfigure[\label{tabA4k}Discrete indices of $\m A_4 \subset SU(3)$.]
{
~\begin{tabular}{|c|r|r|}
\hline ~ && \\[-3mm]
{\begin{tabular}{c} $\m A_4$ \\[1mm]  \,irreps\, \end{tabular}} 
& $\!\begin{array}{c} \wt \ell({\bf r}) \\[1mm] \,(N_\ell\,=\,12) \end{array}\!~$ 
&  $\!\begin{array}{c} \wt A({\bf r})\\[1mm]~(N_A\,=\,2)~ \end{array}\!~$ 
\\[-3mm]~&& \\\hline 
~&&\\[-3mm]
${\bf 1}~$       &      0~~~~~~   &           0~~~~~~ \\
${\bf 1'}$      &  $x$~~~~~~   &        $y$~~~~~~ \\
$\ol{\bf 1'}$       &  $4-x$~~~~~~   &        $-y$~~~~~~ \\
${\bf 3}~$     &     $1$~~~~~~   &          $1$~~~~~~ \\[1mm]\hline
\end{tabular}
}\quad
\subfigure[\label{tabD5k}Discrete indices of $\m D_5 \subset SU(3)$.]
{
\begin{tabular}{|c|r|r|}
\hline ~ && \\[-3mm]
{\begin{tabular}{c}  $\m D_5$ \\[1mm]  \,irreps\,\end{tabular}} 
& $\!\begin{array}{c} \wt \ell({\bf r}) \\[1mm] \,(N_\ell\,=\,20) \end{array}\!~$ 
&  $\!\begin{array}{c} \wt A({\bf r})\\[1mm](N_A\,=\,2) \end{array}\!~$ 
\\[-3mm]~&& \\\hline 
~&&\\[-3mm]
${\bf 1}~$      &      $\!0$~~~~~~~               & 0~~~~~~~ \\
${\bf 1'}$      &  $\!({5\!+\!\xi N_\ell})/{2}$    & $({1}\!+\!\zeta N_A)/{2}$ \\
${\bf 2_1}$     &  $\!(-{3}\!+\!\xi N_\ell)/{2}$   & $({1}\!+\!\zeta N_A)/{2}$ \\
${\bf 2_2}$     &  $\!({13}\!+\!\xi N_\ell)/{2}$   & $({1}\!+\!\zeta N_A)/{2}$ \\[1mm]\hline
\end{tabular}  
}\\
\subfigure[\label{tabS3k}Discrete indices of $\m S_3 \subset SU(3)$.]
{
~\begin{tabular}{|c|r|r|}
\hline ~ && \\[-3mm]
{\begin{tabular}{c} $\m S_3$ \\[1mm]  \,irreps\,\end{tabular}} 
& $\!\begin{array}{c} \wt \ell({\bf r}) \\[1mm] \,(N_\ell\,=\,12) \end{array}\!~$ 
&  $\!\begin{array}{c} \wt A({\bf r})\\[1mm]~(N_A\,=\,2)~ \end{array}\!~$ 
\\[-3mm]~&& \\\hline 
~&&\\[-3mm]
${\bf 1}~$      &     $\!0$~~~~~~~   &           0~~~~~~~ \\
${\bf 1'}$      &  $\!(-{3}\!+\!\xi N_\ell)/{2}$  &  $({1}\!+\!\zeta N_A)/{2}$ \\
${\bf 2}\:$     &  $\!({5}\!+\!\xi N_\ell)/{2}$   &  $({1}\!+\!\zeta N_A)/{2}$ \\[1mm]\hline
\end{tabular}
}
\caption{\label{tabSUM}The definition of the discrete quadratic and cubic
  indices of various 
  finite subgroups of $SU(3)$, namely $\Delta(27)$, $\m S_4$, $\m A_4$, $\m
  D_5$, and $\m S_3$. As some irreps of the finite groups do not occur
  independently from other irreps in the decomposition of $SU(3)$ irreps, the
  definitions of the discrete indices are not always unique. Where present, this
  ambiguity is parameterized by $x_k$, $y_k$; $x$, $y$; $\xi=0,1$, and
  $\zeta=0,1$, respectively.} 
\end{table}


\newpage

\section{\label{so3sect}$\bs{\m S_4}$, $\bs{\m A_4}$, $\bs{\m D_5}$, and $\bs{\m S_3}$ as Subgroups of $\bs{SO(3)}$}
\cleqn
So far, we have considered $\m G$ to be the remnant of a high-energy $SU(3)$
family symmetry. In fact, this is the only possibility for the finite groups
$\m{PSL}_2(7)$, $\m Z_7 \rtimes \m Z_3$, and $\Delta(27)$. On the other hand,
the groups $\m S_4$,  $\m A_4$,  $\m D_5$, and $\m S_3$ can alternatively be
embedded into $SO(3)$.\footnote{A non-trivial embedding into $SU(2)$ is not
  possible since the ${\bf 2}$ as well as the other even-dimensional irreps of
  $SU(2)$ are spinor-like, whereas the irreps of $\m S_4$,  $\m A_4$,  $\m
  D_5$, and $\m S_3$ are not.} Since $SO(3) = SU(2)/\m Z_2$, the indices of
the $SO(3)$ irreps are proportional to the indices of the odd-dimensional
irreps of $SU(2)$. For SU(2), cubic indices are absent, and the quadratic
indices are defined analogously to the $SU(3)$ case
\beq
\mathrm{Trace} \left( \left\{ \, T^{[{\bs \rho}]}_a \,,\,T^{[{\bs \rho}]}_b\,
  \right\}\right) ~=~ \ell({\bs \rho}) \, \delta_{ab} \ .
\eeq
Choosing $T^{[{\bf 2}]}_a =   \sigma_a/2$, with $\sigma_a$
denoting the Pauli matrices, the quadratic index of the fundamental
irrep~${\bf 2}$ is normalized to one. The quadratic indices of all higher 
irreps~${\bs \rho}$ of~$SU(2)$ can then be obtained recursively from
\beq
\ell({\bs \rho} \otimes {\bf 2}) ~=~d({\bs \rho})\,\ell({\bf 2})
\,+\,\ell({\bs \rho}) \,d({\bf 2}) ~=~d({\bs \rho}) \,+\,2\,\ell({\bs \rho}) \ .
\eeq
The indices of the odd-dimensional irreps turn out to be multiples of four.
Hence, a change of normalization yields the following
quadratic indices for the smallest irreps of~$SO(3)$.

{\small
\begin{center}
\begin{tabular}{|c|r|}
\hline  & \\[-3mm]
Irreps ${\bs \rho}$ of $SO(3)$ & $\ell({\bs \rho})$  \\[-3mm] 
& \\ \hline & \\[-3mm]
{\bf 3}        &   1~    \\
{\bf 5}       &   5~    \\
{\bf 7}       &  14~    \\
{\bf 9}       &  30~    \\
{\bf 11}\,      &  55~    \\[1mm]\hline
\end{tabular}
\end{center}
}

\vspace{0mm}

In order to define the discrete indices for the irreps ${\bf r_i}$ of $\m
G \subset SO(3)$, we must first determine how the irreps ${\bs \rho}$
decompose. For the smallest $SO(3)$ irreps, the results are summarized in the
following table. 

{\small
\begin{center}
\begin{tabular}{|c|c|c|c|c|}
\hline  & &&&\\[-3mm]
$SO(3)$ & $\m S_4$ & $\m A_4$ & $\m D_5$ &  $\m S_3$ \\[-3mm] 
& &&&\\ \hline &&&& \\[-3mm]
{\bf 3}  & ${\bf 3_2}$  & ${\bf 3}$  & ${\bf 1'}+{\bf 2_1}$  & ${\bf 1'}+{\bf 2}$     \\
{\bf 5}  & ${\bf 2}+{\bf 3_1}$ & ${\bf 1'}+\ol{\bf 1'}+{\bf 3}$ & ${\bf 1}+{\bf 2_1}+{\bf 2_2}$  & ${\bf 1}+2\!\cdot\!{\bf 2}$    \\
{\bf 7}  & ${\bf 1'}+{\bf 3_1}+{\bf 3_2}$ & ${\bf 1}+2\!\cdot\!{\bf 3}$ & ${\bf 1'}+{\bf 2_1}+2\!\cdot\!{\bf 2_2}$ & ${\bf 1}+2\!\cdot\!({\bf 1'}+{\bf 2})$   \\
{\bf 9}  & ${\bf 1}+{\bf 2}+{\bf 3_1}+{\bf 3_2}$ & ${\bf 1}+{\bf 1'}+\ol{\bf 1'}+2\!\cdot\!{\bf 3}$ & ${\bf 1}+2\!\cdot\!({\bf 2_1}+{\bf 2_2})$ & $2\!\cdot\!{\bf 1}+{\bf 1'}+3\!\cdot\!{\bf 2}$   \\
{\bf 11}\,&${\bf 2}+{\bf 3_1}+2\!\cdot\!{\bf 3_2}$ & ${\bf 1'}+\ol{\bf 1'}+3\!\cdot\!{\bf 3}$  & ${\bf 1}+2\!\cdot\!({\bf 1'}+{\bf 2_1}+{\bf 2_2})$& ${\bf 1}+2\!\cdot\!{\bf 1'}+4\!\cdot\!{\bf 2}$      \\[1mm]\hline
\end{tabular}
\end{center}
}

\noindent Regarding $\m D_5$ there exists an alternative
embedding with the irreps ${\bf 2_1}$ and ${\bf 2_2}$ interchanged. Similar to
the case of $SU(3)$, it is easy to show that the irreps ${\bs \rho}$ of
$SO(3)$ decompose into irreps ${\bf r_i}$ of $\m G$ with the multiplicities
$a_i$ constrained by linear relation. It turns our that these are identical to
the relations we obtained when embedding $\m G$ into $SU(3)$.
\begin{itemize}
\item $\m S_4$: The ${\bf 1'}$ as well as the ${\bf 2}$ are always accompanied
  by an irrep ${\bf 3_1}$.
\item $\m A_4$: The ${\bf 1'}$ and the $\ol{\bf 1'}$ always come in pairs.
\item  $\m D_5$: The sum of the multiplicities of ${\bf 1'}$, ${\bf 2_1}$, and ${\bf 2_2}$ is even.
\item $\m S_3$: The sum of the multiplicities of ${\bf 1'}$ and ${\bf 2}$ is even.
\end{itemize}
Therefore, the discrete indices are not defined uniquely. From the
decompositions of the smallest $SO(3)$ irreps we can determine $\wt \ell({\bf
  r_i})$, with the arbitrariness parameterized by $x$ and $\xi=0,1$,
respectively. 
\begin{itemize}
\item $\m S_4$: First we set $\wt \ell({\bf 3_2})=1$. Introducing the
  parameter $x=\wt \ell({\bf 3_1})$, we find from the decomposition of the
  ${\bf 5}$ that $\wt \ell({\bf 2})=5-x$. Similarly, the ${\bf 7}$ fixes $\wt
  \ell({\bf 1'})=13-x$. Inserting these values for the discrete indices into
  the decomposition of the ${\bf 9}$, we see that the quadratic indices can
  only be defined modulo $N_\ell=24$.
\item  $\m A_4$: Here we have $\wt \ell({\bf 3})=1$. Defining $x=\wt \ell({\bf
    1'})$, the decomposition of the ${\bf 5}$ yields $\wt \ell(\ol{\bf
    1'})=4-x$. With these values, one finds that $N_\ell=12$.
\item $\m D_5$: Comparing the decomposition of the ${\bf 5}$ and the ${\bf 9}$
  shows that $N_\ell=20$. The value for $2\, \wt \ell({\bf
    2_2})=13~\mathrm{mod}~N_\ell$ is obtained from combining the $SO(3)$ irreps
  ${\bf 3}$ and~${\bf 7}$. With $\xi=0,1$ this gives $\wt \ell({\bf
    2_2})=(13+\xi N_\ell)/2$. Then the ${\bf 5}$ fixes $\wt \ell({\bf
    2_1})=(-3+\xi N_\ell)/2$. Finally, from the ${\bf 3}$ we get $\wt \ell({\bf
    1'})=(5+\xi N_\ell)/2$.
\item $\m S_3$: Comparison of the $SO(3)$ irreps ${\bf 3}$ and ${\bf 7}$
  yields $N_\ell=12$. From the ${\bf 5}$ we determine $\wt \ell({\bf 2})=(5+\xi
  N_\ell)/2$. Then we find $\wt \ell({\bf 1'})=(-3+\xi N_\ell)/2$ from the
  ${\bf 3}$. 
\end{itemize} 
Remarkably, all these quadratic indices are identical to the results
obtained in Section~\ref{othersect} where $\m S_4$,  $\m A_4$,  $\m D_5$, and  
$\m S_3$ are considered to be subgroups of $SU(3)$. Concerning the quadratic
indices it therefore does not make any difference at all whether $\m G$
originates in $SU(3)$ or $SO(3)$. Of course, in the latter case the cubic
indices are absent.

\section{\label{DACsect}Discrete Anomaly Conditions}
\cleqn
Having defined the discrete indices for the finite groups $\m{PSL}_2(7)$, $\m
Z_7 \rtimes \m Z_3$, $\Delta(27)$, $\m S_4$, $\m A_4$, $\m D_5$, and $\m S_3$,
we are now in a position to formulate the corresponding discrete anomaly
conditions. Our starting point is to require anomaly freedom of the
underlying continuous family symmetry $G_f$. Under the assumption that
$G_f=SU(3)$, we obtain two anomaly cancellation conditions 
\beq
\sum_k \ell_k Y_k ~=~ 0 \ , \qquad  \sum_k A_k ~=~ 0 \ .\notag
\eeq
Here $k$ labels the fermions of the complete theory, with $\ell_k$ and $A_k$
being the quadratic and the cubic index corresponding to the particle's
$SU(3)$ irrep. $Y_k$ denotes the hypercharge in the normalization where the
left-handed quark doublet has $Y_Q = 1$. In the following, we further
assume that no particle $k$ has fractional hypercharge in this
normalization.\footnote{Particles with fractional hypercharge in this
normalization are electrically charged. Furthermore, they cannot decay into
Standard Model particles alone, so that the lightest such particle would be
stable. Since dark matter should be neutral, their existence is disfavored.}
After the breakdown of $SU(3)$  to the non-Abelian finite symmetry $\m G$, the
$SU(3)$ irreps decompose into irreps of $\m G$. Labeling these by $i$, the
discrete anomaly cancellation conditions can be obtained from
\beqn
\sum_{i = \mathrm{light}} \wt \ell_i \, Y_i ~~ + \sum_{i = \mathrm{massive}}
\wt \ell_i \, Y_i  &=& 0 \mathrm{~mod~}N_\ell \ , \label{SSU}\\
\sum_{i = \mathrm{light}} \wt A_i ~~ + \sum_{i = \mathrm{massive}}
\wt A_i   &=& 0 \mathrm{~mod~}N_A \ ,\label{SSS}
\eeqn
with $N_I$ depending on the specific group $\m G$. 
For $G_f=SU(2)~\mathrm{or}~SO(3)$, the cubic anomaly does not exist so
that we are left with Eq.~(\ref{SSU}) only. In the following, we evaluate
the sums over the massive degrees of freedom in Eqs.~(\ref{SSU}) and
(\ref{SSS}), showing that they can be incorporated into the right-hand
side, in some cases changing the value of $N_I$. Thus we are lead to the
discrete anomaly conditions which constrain the irreps of $\m G$ assigned to
the light fermions.

\subsection{Mass Terms and Their Effects on the Anomaly Conditions}
Before elaborating on the non-Abelian case, it might be useful to remind
ourselves of how massive fermions enter the anomaly equations in a scenario
where the discrete symmetry is~$\m Z_N$
\cite{Ibanez:1991hv,Ibanez:1991pr}. Such an Abelian discrete symmetry arises
when a $U(1)$ gauge symmetry gets spontaneously broken by the vacuum
expectation value (VEV) of a SM singlet field with $U(1)$ charge $N$. As a
result of this breaking, some fermions, the so-called massive fermions,
will have a bilinear mass term whose $U(1)$ charge is an integer multiple
of~$N$. Using the standard conventions, a pair of massive fermionic fields can
only contribute 
an integer multiple of $N/2$ to the anomaly equations. Therefore the discrete
anomaly conditions are  given modulo~$N$, with $N$ directly related to the
spontaneous breaking of the continuous $U(1)$ symmetry. 

The situation becomes more involved with non-Abelian symmetries for two
reasons. First, the details of how $G_f$ breaks down to $\m G$ are ambiguous
as can be seen, e.g., from the decomposition of $SU(3)$ irreps into irreps of
$\m Z_7 \rtimes \m Z_3$ [cf. above Eq.~(\ref{together1})]. Even when
restricting to the irreps of $SU(3)$ up to ${\bf 27}$, there are six irreps
which can acquire a VEV (in a suitable direction) and leave the discrete
symmetry $\m Z_7 \rtimes \m Z_3$ unbroken. Instead of being related to the
breaking of the continuous to the discrete family symmetry, the values
for~$N_I$ in the modulo~$N_I$ of the  discrete anomaly equations mainly
originate from the definition of the discrete indices. Second, the
possibilities of forming bilinear mass terms in the presence of a non-Abelian
discrete symmetry are constrained by the Kronecker products. Particles which
acquire a mass at the breaking of $G_f$ must have a $\m G$ invariant bilinear
mass term since the $SU(3)$ irreps which are chosen to break the family gauge
symmetry can only get a VEV in the direction that singles out the singlet
${\bf 1}$ of $\m G$. In order to discuss the effects of the massive fermionic
fields on the anomaly conditions, it is therefore necessary to determine those
Kronecker products which contain a singlet of the finite group.

There are two types of massive fermionic fields, Majorana particles and Dirac
particles. As Majorana particles are necessarily neutral under any $U(1)$ they
don't contribute to Eq.~(\ref{SSU}). 
On the other hand, the Dirac degrees of freedom
always come in pairs with the two fields having opposite hypercharge; therefore
their contribution to Eq.~(\ref{SSU}) is
$$
\wt \ell_{i_1} \, Y_{i_1}  \,+\,\wt \ell_{i_2} \, Y_{i_2} ~=~( \wt \ell_{i_1}\,
-\,\wt \ell_{i_2}) \, Y_{i_1} \ .
$$
Below we evaluate this term as well as the contribution of the massive fields
to Eq.~(\ref{SSS}) explicitly for the various finite groups in turn.
\begin{itemize}
\item {$\bf {\m{PSL}_2(7)}$:} The Kronecker products~\cite{Luhn:2007yr} show
that we obtain invariant bilinear terms from the products ${\bf 3} \otimes
\ol{\bf   3}$, ${\bf 6} \otimes {\bf 6}$,  ${\bf 7} \otimes {\bf 7}$,  ${\bf 8}
\otimes {\bf 8}$. Applying the discrete indices defined in
Table~\ref{tabPSL}(a), there is no contribution of massive fields to
Eq.~(\ref{SSU}), and only the sextet yields a non-zero contribution to
Eq.~(\ref{SSS}). As the ${\bf 6}$ is a real representation of $\m{PSL}_2(7)$,
it can correspond to a Majorana field. In that case, because $\wt
A_{\bf 6} = 7 $, the value of $N_A$ on the right-hand side of
Eq.~(\ref{SSS}) is reduced from 14 to 7. Hence we have
\beq
\sum_{i = \mathrm{light}} \wt \ell_i \, Y_i ~=~ 0 \mathrm{~mod~} 24\ , \qquad
\sum_{i = \mathrm{light}} \wt A_i    ~=~ 0 \mathrm{~mod~}7 \ .\label{PSL-Acond}
\eeq

\item {$\bf {\m Z_7 \rtimes \m Z_3}$:} The bilinear group invariants are
obtained from ${\bf 1'} \otimes \ol{\bf 1'}$ and ${\bf 3} \otimes \ol{\bf 3}$,
which shows that no massive Majorana particles are allowed. Only a Dirac pair
of the former type has a contribution to a discrete anomaly condition which
does not automatically vanish. From Table~\ref{tabPSL}(b)
we find that such a Dirac pair adds $(2x-1)\,Y_i$ in Eq.~(\ref{SSU}), which
vanishes modulo~3 for $$x~=~\frac{1}{2}~\mathrm{or}~2\ .$$ The massive Dirac
pair then does not contribute to Eq.~(\ref{SSU}) at all. This yields
 \beq
\sum_{i = \mathrm{light}} \wt \ell_i \, Y_i ~=~ 0 \mathrm{~mod~} 3\ , \qquad
\sum_{i = \mathrm{light}} \wt A_i    ~=~ 0 \mathrm{~mod~}7 \ .\label{Z7Z3-Acond}
\eeq

\item {$\bf { \Delta(27)}$:} Here the bilinear mass terms stem from the
  products ${\bf 1_{2l-1}} \otimes {\bf 1_{2l}}$ with $l=1,...,4$ as well as
  ${\bf 3} \otimes \ol{\bf 3}$. Again no massive Majorana particles are
  possible. Choosing 
\beq
x^{}_{2l-1} ~=~x^{}_{2l}  \quad \mathrm{with} \quad \sum_{l=1}^4 x^{}_{2l} ~=~ 0\ ,\qquad \mathrm{~~and~~} \qquad 
y^{}_{2l-1} ~=~ - \, y^{}_{2l} \ ,      \notag
\eeq
for the discrete indices of Table~\ref{tabSUM}(a), the massive Dirac particles
drop out of Eqs.~(\ref{SSU}) and (\ref{SSS}), yielding
 \beq
\sum_{i = \mathrm{light}} \wt \ell_i \, Y_i ~=~ 0 \mathrm{~mod~} 3\ , \qquad
\sum_{i = \mathrm{light}} \wt A_i    ~=~ 0 \mathrm{~mod~}9 \ .\label{Delta-Acond}
\eeq

\item {$\bf {\m S_4}$:} As mass terms can be built from ${\bf 1'} \otimes
  {\bf 1'}$, ${\bf 2} \otimes {\bf 2}$, and ${\bf 3_i} \otimes {\bf 3_i}$
massive particles can be of both Majorana as well as Dirac type. Regardless of
the value of $x$ they do not contribute to Eq.~(\ref{SSU}). On the other hand,
a heavy Majorana particle living in the irrep ${\bf 3_2}$ adds $\wt A_{\bf
  3_2}=1$ to Eq.~(\ref{SSS}), therefore changing $N_A$ from 2 to 1 on the
right-hand side. Taking $$y~=~0\mathrm{~or~}1 \ ,$$ all discrete cubic indices
are integer, leading to no useful constraint on the light particle spectrum from
Eq.~(\ref{SSS}). Using the discrete indices listed in Table~\ref{tabSUM}(b)
with arbitrary $x$, a non-trivial condition only results from Eq.~(\ref{SSU})
which reads 
\beq
\sum_{i = \mathrm{light}} \wt \ell_i \, Y_i ~=~ 0 \mathrm{~mod~} 24\ .\label{S4-Acond}
\eeq

\item {$\bf {\m A_4}$:} The bilinear invariants are obtained from the products
  ${\bf 1'} \otimes \ol{\bf 1'}$ and ${\bf 3} \otimes {\bf
    3}$. For $$x~=~2~\mathrm{or}~8\ ,$$ massive particles do not
  contribute to Eq.~(\ref{SSU}). The possibility of having a heavy Majorana
  particle in the irrep ${\bf 3}$ reduces $N_A$ from 2 to 1 on the right-hand
  side of Eq.~(\ref{SSS}). Irrespective of the value for $y$, a Dirac pair
  with the mass term ${\bf 1'} \otimes \ol{\bf 1'}$ does not contribute to
  Eq.~(\ref{SSS}). Therefore the resulting discrete anomaly condition is
  non-trivial; it is equivalent to the requirement of having as many
  ${\bf 1'}$ as there are~$\ol{\bf 1'}$ in the light particle
  content. Explicitly, with the indices defined in
  Table~\ref{tabSUM}(c), the two discrete anomaly conditions are given as 
\beq
\sum_{i = \mathrm{light}} \wt \ell_i \, Y_i ~=~ 0 \mathrm{~mod~} 12\ , \qquad
\sum_{i = \mathrm{light}} \wt A_i    ~=~ 0 \mathrm{~mod~}1 \ .\label{A4-Acond}
\eeq

\item {$\bf {\m D_5}$:} The masses of heavy particles can be generated from
  the products ${\bf 1'} \otimes {\bf 1'}$ and ${\bf 2_i} \otimes {\bf 2_i}$.
Such fields can be of either Majorana or Dirac type. They do not contribute to
Eq.~(\ref{SSU}). However, each Dirac particle adds an odd integer to
Eq.~(\ref{SSS}); allowing for heavy Majorana particles, we obtain
half-odd integer contributions. Since the discrete cubic indices
are multiples of $\frac{1}{2}$, no useful constraint is obtained from
Eq.~(\ref{SSS}). With the indices given in Table~\ref{tabSUM}(d), the
discrete anomaly condition obtained from Eq.~(\ref{SSU}) yields
\beq
\sum_{i = \mathrm{light}} \wt \ell_i \, Y_i ~=~ 0 \mathrm{~mod~} 20\ .\label{D5-Acond}
\eeq

\item {$\bf {\m S_3}$:} The bilinear invariants can originate from ${\bf 1'}
  \otimes {\bf 1'}$ and ${\bf 2} \otimes {\bf 2}$. As in the case of the group
  $\m D_5$, massive particles do not contribute to Eq.~(\ref{SSU}), while
  their possible existence renders Eq.~(\ref{SSS}) useless. With the 
  indices shown in Table~\ref{tabSUM}(e), the non-trivial discrete anomaly
  condition reads 
\beq
\sum_{i = \mathrm{light}} \wt \ell_i \, Y_i ~=~ 0 \mathrm{~mod~} 12\ .\label{S3-Acond}
\eeq

\end{itemize}

Eqs.~(\ref{PSL-Acond})-(\ref{S3-Acond}) show that the discrete anomaly
conditions on the light particle spectrum depend on the finite group~$\m G$.
All are subgroups of $SU(3)$, however, $\m S_4$, $\m A_4$, $\m D_5$, and $\m
S_3$ can alternatively be embedded into $SO(3)$. As the discrete quadratic
indices are identical for $SU(3)$ and $SO(3)$, the resulting discrete $G_f -
G_f - U(1)_Y$  anomaly conditions are identical too. Of course, for an
embedding into $SO(3)$ no cubic anomaly exists. Interestingly, even an $SU(3)$
origin of the finite groups $\m S_4$, $\m D_5$, and $\m S_3$ does not yield a
discrete cubic anomaly condition. Only in the case of $\m A_4$, the second
condition of Eq.~(\ref{A4-Acond}) is rendered useless because $\m A_4$ could
originate in $SO(3)$ instead of $SU(3)$. For the sake of quick reference for
flavor model builders, we summarize the discrete anomaly conditions together
with the necessary discrete indices in Appendix~\ref{mainresults}.

We emphasize that they are obtained under the assumption of an underlying
anomaly-free gauge symmetry $G_f$. 
In the case where $G_f=U(1)$, it can be argued that the
resulting {\it linear} discrete anomaly conditions are identical to the
requirement that the effective instanton vertex for the SM gauge theory
respect the remnant $\m Z_N$~symmetry~\cite{Banks:1991xj}. However,
non-perturbative effects cannot explain the condition arising from the cubic
anomaly $U(1)-U(1)-U(1)$, although it carries interesting information about the
necessity of fractionally charged particles. Similarly, the instanton argument
can also be applied to non-Abelian discrete symmetries (see
e.g. Ref.~\cite{Babu:2007mb}). It is however beyond the scope of this paper to
investigate the relations between the discrete anomaly conditions of
Eqs.~(\ref{PSL-Acond})-(\ref{S3-Acond}) and the constraints arising from the
requirement that the non-perturbative processes be invariant under the
respective discrete symmetry.

In the following section we apply our anomaly conditions to some existing
flavor models to see whether or not their preferred non-Abelian finite
symmetry can be a remnant of $SU(3)$ or $SO(3)$, respectively.




\section{\label{casessect}Case Studies}
\cleqn

In order to illustrate the use of our work we examine some existing models of
flavor. Since the SM quarks and leptons belong to the light
particle spectrum, the discrete anomaly conditions can only be
evaluated if the assignment of these fermions to irreps of the finite group
$\m G$ is completely given. Depending on the model, the right-handed 
neutrinos~$\nu^c$ might also remain massless after $G_f$ is broken down to $\m
G$. Particularly in supersymmetric models one encounters additional fermionic 
degrees of freedom which in general may contribute to the discrete
anomalies. Here, we restrict our study to models where this is not the case,
i.e. only the SM fermions (possibly including $\nu^c$) contribute to the
discrete anomaly equations, whereas other fermions are either absent,
transform trivially under $\m G$, or have $\m G$-invariant bilinear mass
terms.\footnote{For $\m G \subset SO(3)$, light fermions with zero hypercharge
  do not enter the anomaly equation either.}

We first note that the sum of the hypercharges of all quarks and leptons is
zero, i.e. the SM does not have a $\mathrm{Gravity}-\mathrm{Gravity}-U(1)_Y$
anomaly. Therefore the mixed discrete anomaly $\m G-\m G-U(1)_Y$ vanishes
identically if the SM fermions all live in the same representation of the
finite group $\m G$. This is the case for the models of
Refs.~\cite{Luhn:2007sy,deMedeirosVarzielas:2006fc,Hagedorn:2006ug,Morisi:2007ft}. The discrete symmetries employed 
in~\cite{Luhn:2007sy,deMedeirosVarzielas:2006fc} 
are $\m Z_7\rtimes \m Z_3$ and $\Delta(27)$, respectively, which can
only be embedded in $SU(3)$. One therefore still has to check the discrete cubic
anomaly. With the fermions transforming as triplets of $\m G$ in both cases,
we have\\
$$
\sum_{i=\mathrm{light}} \wt A_i ~=~ \sum_{i=\mathrm{light}} 1 ~=~16 \ .
$$\\
The comparison with Eqs.~(\ref{Z7Z3-Acond}) and~(\ref{Delta-Acond}) shows that
both models are not discrete anomaly free and therefore incomplete: they
require additional fermions which do not acquire mass when $SU(3)$ is broken
down to the discrete family symmetry $\m G$. 

As mentioned above, the models of Refs.~\cite{Hagedorn:2006ug,Morisi:2007ft}
have no mixed discrete anomaly $\m G-\m G-U(1)_Y$. Since the applied family
symmetries $\m S_4$ and $\m A_4$ are subgroups of $SO(3)$, these models are
not constrained by the cubic anomaly condition and therefore discrete anomaly
free. Similarly, one has to check only the mixed discrete anomaly for the
examples listed in Table~\ref{tabcasestudy}. In all models, the SM fermions
are the only particles contributing to the discrete anomaly, which is
therefore determined solely by the assignment of the quarks and leptons to
irreps of $\m G$. Whenever the value for the discrete anomaly, given in the
rightmost column, is non-zero, the model is not discrete anomaly free,
i.e. it is necessary to include additional light fermions or, alternatively, 
heavy fermions with fractional hypercharges.

\begin{table}\begin{center}
$
\begin{array}{||c|c||c|c|c|c|c||c||} \hline\hline &&&&&&&\\[-3mm]
\mathrm{Group} & \mathrm{Refs.}& Q & u^c & d^c & L & e^c 
& \sum \wt \ell_i  Y_i \\[2mm] \hline\hline &&&&&&&\\[-3mm]
\m A_4 & \mbox{\cite{Ma:2001dn,Babu:2002dz,Altarelli:2005yx}}& {\bf 3} & {\bf 1},{\bf
  1'},{\ol{\bf 1'}} & {\bf 1},{\bf 1'},{\ol{\bf 1'}} &  {\bf 3} & {\bf 1},{\bf 1'},{\ol{\bf 1'}} & 0\mathrm{\,mod\,}12  \\[1.7mm] \hline &&&&&&&\\[-3mm]
\m A_4 &\mbox{\cite{Ma:2001dn}}& {\bf 3} & {\bf 3} & {\bf 3} &  {\bf 3} & {\bf 1},{\bf 1'},{\ol{\bf 1'}} & 6\mathrm{\,mod\,}12  \\[1.7mm] \hline &&&&&&&\\[-3mm]
\m A_4 & \mbox{\cite{Ma:2001dn,Babu:2005se,Altarelli:2005yx}}& {\bf 1},{\bf 1},{\bf 1} & {\bf 1},{\bf 1},{\bf 1} & {\bf 1},{\bf 1},{\bf 1} &  {\bf 3} & {\bf 1},{\bf 1'},{\ol{\bf 1'}} & 6\mathrm{\,mod\,}12  \\[1.7mm] \hline &&&&&&&\\[-3mm]
\m A_4 & \mbox{\cite{King:2006np}}& {\bf 3} & {\bf 1},{\bf 1},{\bf 1} & {\bf 1},{\bf 1},{\bf 1} &  {\bf 3} & {\bf 1},{\bf 1},{\bf 1} & 0\mathrm{\,mod\,}12  \\[1.7mm] \hline &&&&&&&\\[-3mm]
\m A_4 & \mbox{\cite{Hirsch:2007kh}}& {\bf 1},{\bf 1},{\bf 1} & {\bf 1},{\bf 1},{\bf 1} & {\bf 1},{\bf 1},{\bf 1} &  {\bf 1},{\bf 1'},{\ol{\bf 1'}} & {\bf 3} & 6\mathrm{\,mod\,}12  \\[1.7mm] \hline &&&&&&&\\[-3mm]
\m D_5 & \mbox{\cite{Hagedorn:2006ir}} &  {\bf 1},{\bf 2_2} & {\bf 1},{\bf 2_1} & {\bf 1},{\bf 2_1} &  {\bf 1},{\bf 2_2} &{\bf 1},{\bf 2_1}  & 0\mathrm{\,mod\,}20  \\[1.7mm] \hline &&&&&&&\\[-3mm]
\m S_3 & \mbox{\cite{Babu:2007zm}} & {\bf 1},{\bf 1},{\bf 1} & {\bf 1},{\bf 1},{\bf
  1} & {\bf 1},{\bf 1},{\bf 1} &  {\bf 1},{\bf 2} &{\bf 1},{\bf 2}  & 0\mathrm{\,mod\,}12  \\[1.7mm] \hline &&&&&&&\\[-3mm]
\m S_3 & \mbox{\cite{Feruglio:2007hi}} & {\bf 1},{\bf 2} & {\bf 1},{\bf 1},{\bf
  1} & {\bf 1},{\bf 1},{\bf 1} &  {\bf 1},{\bf 2} &{\bf 1'},{\bf 1'},{\bf 1'}
& 3\mathrm{\,mod\,}12  \\[1.7mm] \hline \hline 
\end{array}
$
\caption{\label{tabcasestudy}With the particle content given in these existing
  flavor models, only the SM fermions contribute to the mixed discrete
  anomaly. Thus the assignment of the quarks and leptons to irreps of a finite
  group determines whether a model is discrete anomaly~free.}
\end{center} \end{table}

\newpage

\section{\label{conclsect}Conclusion}
\cleqn
In recent years, many flavor models invoking the operation of a 
non-Abelian discrete family symmetry have been suggested to explain the
tri-bimaximal leptonic mixing pattern. This plethora of possibilities asks for
criteria to assess the viability of a model. In our study we have formulated
the consequences of embedding non-Abelian discrete symmetries $\m G$ into a
continuous gauge symmetry $G_f$. Mathematical consistency requires the
underlying gauge theory to be anomaly free; this translates to  discrete
anomaly conditions after the breaking of $G_f$. A model builder's toolbox for
quickly checking the discrete anomaly conditions of a model is provided for in
Appendix~\ref{mainresults}.

\section*{Acknowledgments}
We are indebted to G.~Ross for stimulating discussions during the initial
stages of this~work. We also wish to thank Y.~Tachikawa for his inductive
proof for the cubic index of $\m{PSL}_2(7)$. PR acknowledges support
from the Institute for Advanced Study, the Ambrose Monell foundation, and 
the Department of Energy Grant No. DE-FG02-97ER41029. CL is supported by
the University of Florida through the Institute for Fundamental Theory.

\vspace{1mm}

\begin{appendix}
\section{\label{young}Obtaining all $\bs{SU(3)}$ Representations Successively}
\cleqn
It is well known that the irreps of $SU(3)$ can be constructed from solely the
fundamental triplet. Still, in the inductive step of our proof of
Eq.~(\ref{consistent}) we check the validity of this equation not only for
${\bs \rho} \otimes {\bf 3}$ but also for ${\bs \rho} \otimes \ol{\bf 3}$. Considering the
$\ol{\bf 3}$ as well can potentially add new constraints to the definition of
the discrete indices. For example, without the $\ol{\bf 3}$, 
the first table of Appendix~\ref{f-proof} suggests that the discrete cubic
indices of $\m{PSL}_2(7)$ could be defined modulo $N_A=28$; however, the lower
half of the table reveals that, in fact, we need $N_A=14$. 

To better understand the reason for why the $\ol{\bf 3}$ must be included in
our proof, let us discuss all possible ways for obtaining the ${\bf 15}$ of
$SU(3)$ by multiplying the ${\bf 3}$ with some smaller irrep for which
Eq.~(\ref{consistent}) shall hold. From the Young
tableau of the ${\bf 15}$\\
$$\Yinterspace{4mm} \yng(3,1)  ,
$$\\
it is obvious that there are only two smaller irreps which, after
multiplication with the~${\bf 3}$, include the ${\bf 15}$. The corresponding
products are \\
\beqn 
\Yinterspace{4mm} \yng(2,1) \otimes \yng(1) 
&=& \Yinterspace{4mm} \yng(3,1) \oplus \yng(2,2) \oplus \yng(1)  , 
\notag \\[3mm]
\Yinterspace{4mm} \yng(3) \otimes \yng(1) 
&=& \Yinterspace{4mm} \yng(3,1) \oplus \yng(4)   . \notag
\eeqn\\
In both cases, we obtain the ${\bf 15}$ {\it and another  new} irrep for which
the validity of Eq.~(\ref{consistent}) has not been shown.\footnote{One might
have the impression that the $\ol{\bf 6}$ in the first product should
{\it automatically} satisfy Eq.~(\ref{consistent}). This however is incorrect
since, for the cubic index, we cannot infer that $-A({\bf 6}) = A(\ol{\bf
  6})=\wt A({\bf  6})~\mathrm{mod~}N_A$ from the validity of $A({\bf 6}) = \wt
A({\bf   6})~\mathrm{mod~}N_A$ only. 
One must prove Eq.~(\ref{consistent}) separately for the $\ol{\bf 6}$.}
Therefore, with multiplications by ${\bs \sigma} ={\bf 3}$, we can only prove that
Eq.~(\ref{consistent}) holds true for {\it a sum of two new} irreps.  

This shortcoming can be overcome by adding the choice of ${\bs \sigma} = \ol{\bf
  3}$. Then, all irreps can be successively generated  with {\it only one new}
irrep  appearing on the right-hand side of the corresponding products. 
Assume that we knew all irreps of the form\\
\beq 
\Yinterspace{4mm} \young(12\cdot\cdot\cdot k) , \quad 
 \young(12\cdot\cdot\cdot k,1) ,\label{you}
\eeq\\
with $1\leq k \leq K$. Notice that the case with $k=1$ comprises the basis of
our proof by induction; hence we must initially show that
Eq.~(\ref{consistent}) is true for both, the ${\bf 3}$ and the $\ol{\bf
  3}$. We now determine the two products \\[-1mm]
\beqn
\Yinterspace{3mm} \young(12\cdot\cdot\cdot K) 
\otimes  \yng(1) 
&\!\!=\!\!&\Yinterspace{3mm}
 \young(12\cdot\cdot\cdot K~)  \oplus
 \young(12\cdot\cdot\cdot K,~)  , ~~~  \label{you1} \\[3mm]
\Yinterspace{3mm} \young(12\cdot\cdot\cdot K) 
\otimes  \yng(1,1) 
&\!\!=\!\!&\Yinterspace{3mm}
 \young(12\cdot\cdot\cdot K~,~)  \oplus
 \young(2\cdot\cdot\cdot K)  . ~~~  \label{you2} 
\eeqn\\
The second irreps on the right-hand sides of Eqs.~(\ref{you1}) and (\ref{you2})
are already known. The first ones are new and extend Eq.~(\ref{you}) to $k\leq
K+1$. This shows that the irreps 
in Eq.~(\ref{you}) can be obtained with arbitrary $k\in \mathbb N$.

We can now fill up the second row of the Young tableaux by multiplications
with ${\bf 3}$. Assume that the irreps of the form \\
\beq
\Yinterspace{4mm} \young(12\cdot\cdot\cdot k,12\cdot l) , \label{you3} 
\eeq\\
are known for arbitrary $k\in \mathbb N$ and $1 \leq l\leq L$. Notice that the
case $L=1$ is nothing but the second Young tableau of Eq.~(\ref{you}). Then\\[-1mm]
\beqn
\Yinterspace{1mm} \young(12\cdot\cdot\cdot k,12\cdot L) \otimes \young(~) 
&\!\!=\!\!& \Yinterspace{1mm}
\young(12\cdot\cdot\cdot k,12\cdot L~) \oplus
\young(12\cdot\cdot\cdot k~,12\cdot L)\oplus
\young(2\cdot\cdot\cdot k,2\cdot L)  \ . \notag
\eeqn\\
Only the first irrep on the right-hand side is a new one and extends
Eq.~(\ref{you3}) to $l\leq L+1$, and thus to arbitrary $l\in \mathbb
N$. Therefore any irrep of $SU(3)$ can be 
generated successively by multiplication with ${\bf 3}$ and $\ol{\bf 3}$ in a
way that only one new irrep occurs on the right-hand side of the tensor
product.\footnote{It is worth mentioning that this method can be generalized
  to Lie groups other than $SU(3)$. Then ${\bs \sigma}$ has to take all
  irreps associated with the fundamental weights: for example, in $SU(4)$ these
  are ${\bf 4}$, ${\bf 6}$, $\ol{\bf 4}$. Thanks to Dr. Yuji Tachikawa
  for pointing this out.}

\newpage

\section{\label{f-proof}The Proof of Eq.~(\ref{effs}) for $\bs{\m{PSL}_2(7)}$ and $\bs{\m Z_7 \rtimes \m Z_3}$}
\cleqn

This appendix shows the explicit values of $f_I^{\,i}({\bs \sigma})$ and
$\mathfrak f_I^{\,i}({\bs \sigma})$  for the finite groups $\m{PSL}_2(7)$ and
$\m Z_7 \rtimes \m Z_3$.  They are calculated from
Eqs.~(\ref{l3})-(\ref{rsigma}) with the discrete indices given in
Table~\ref{tabPSL}. The comparison proves that Eq.~(\ref{effs}) is satisfied
for all $i$. Therefore our definition of the discrete indices is consistent.

\vspace{4mm}

\begin{center}
\begin{tabular}{||c||c|c||c|c||} \hline  \hline 
&\multicolumn{2}{|c||}{}&\multicolumn{2}{|c||}{} \\[-2mm]
${\bs{\m{PSL}_2(7)}}$ & \multicolumn{2}{|c||}{Quadratic Index $\ell$ ($N_\ell=24$)} 
& \multicolumn{2}{|c||}{Cubic Index $A$ ($N_A=14$)} \\[2mm] \hline\hline  
&&&& \\[-3mm]
$i$ & $f_\ell^{\,i}({\bf 3})$ & $\mathfrak f_\ell^{\,i}({\bf 3})$ & $f_A^{\,i}({\bf 3})$ & $\mathfrak f_A^{\,i}({\bf 3})$ \\[1mm] \hline
$0$ & $1+0=1$ & $1=1$ & $1+0=1$ & $1=1$  \\ 
$1$ & $3+3=6$ & $1+5=6$ & $3+3=6$ & $-1+7=6$  \\ 
$2$ & $3+3=6$ & $0+6=6$ & $3-3=0$ & $0+0=0$   \\ 
$3$ & $6+15=21$ & $1+14+6=21$ & $6+21=27$ & $-1+0+0=-1$   \\ 
$4$ & $7+42=49$ & $5+14+6=25$ & $7+0=7$ & $7+0+0=7$   \\ 
$5$ & $8+18=26$ & $1+5+14+6=26$ & $8+0=8$ & $1+7+0+0=8$   \\ \hline\hline
&&&& \\[-3mm]
$i$ &  $f_\ell^{\,i}(\ol{\bf 3})$& $\mathfrak f_\ell^{\,i}(\ol{\bf 3})$&  $f_A^{\,i}(\ol{\bf 3})$&$\mathfrak f_A^{\,i}(\ol{\bf 3})$\\[1mm] \hline
$0$ & $1+0=1$ & $1=1$ & $-1+0=-1$ & $-1=-1$  \\ 
$1$ & $3+3=6$ & $0+6=6$ & $-3+3=0$ & $0+0=0$  \\ 
$2$ & $3+3=6$ & $1+5=6$ & $-3-3=-6$ & $1+7=8$   \\ 
$3$ & $6+15=21$ & $1+14+6=21$ & $-6+21=15$ & $1+0+0=1$   \\ 
$4$ & $7+42=49$ & $5+14+6=25$ & $-7+0=-7$ & $7+0+0=7$   \\ 
$5$ & $8+18=26$ & $1+5+14+6=26$ & $-8+0=-8$ & $-1+7+0+0=6$   \\ \hline\hline
\end{tabular}\end{center}

\vspace{2mm}

\begin{center}
\begin{tabular}{||c||c|c||c|c||} \hline  \hline 
&\multicolumn{2}{|c||}{}&\multicolumn{2}{|c||}{} \\[-2mm]
${\bs{\m Z_7 \rtimes \m Z_3}}$ & \multicolumn{2}{|c||}{Quadratic Index $\ell$ ($N_\ell=3$)} 
& \multicolumn{2}{|c||}{Cubic Index $A$ ($N_A=7$)} \\[2mm] \hline\hline  
&&&& \\[-3mm]
$i$ 
& \hspace{6.5mm}$f_\ell^{\,i}({\bf 3})$\hspace{6.5mm} 
& \hspace{6.5mm}$\mathfrak f_\ell^{\,i}({\bf 3})$\hspace{6.5mm} 
& \hspace{6.5mm}$f_A^{\,i}({\bf 3})$\hspace{6.5mm} 
&\hspace{6.5mm} $\mathfrak f_A^{\,i}({\bf 3})$\hspace{6.5mm} \\[1mm] \hline
$0$   & $1$     & $1$ & $1$ & $\phantom{-}1$~  \\ 
$1+2$ & $5$     & $2$ & $2$ & $\phantom{-}2$~  \\ 
$3$   & $6$     & $3$ & $6$ & $-1$~            \\ 
$4$   & $6$     & $3$ & $0$ & $\phantom{-}0$~  \\ \hline\hline
&&&& \\[-3mm]
$i$ &  $f_\ell^{\,i}(\ol{\bf 3})$& $\mathfrak f_\ell^{\,i}(\ol{\bf 3})$&  $f_A^{\,i}(\ol{\bf 3})$&$\mathfrak f_A^{\,i}(\ol{\bf 3})$\\[1mm] \hline
$0$   & $1$ & $1$ & $-1$~           & $-1$~             \\ 
$1+2$ & $5$ & $2$ & $-2$~           & $-2$~             \\ 
$3$   & $6$ & $3$ & $\phantom{-}0$~ & $\phantom{-}0$~   \\ 
$4$   & $6$ & $3$ & $-6$~           & $\phantom{-}1$~   \\ \hline\hline
\end{tabular}\end{center}

\newpage

\section{\label{mainresults}Toolbox for Model Builders}
\cleqn

In this appendix we collect all the results of our study which are relevant
for flavor model building. We tabulate the discrete indices for each finite
group, now taking into account the constraints from the mass terms discussed in
Section~\ref{DACsect}. Thus some of the parameters of Tables~\ref{tabPSL}
and~\ref{tabSUM} are fixed. Others remain undetermined and the discrete
anomaly conditions must be satisfied for arbitrary values. Only in the case of
$\Delta(27)$ the parameters $x_{2l}$ with $l=1,...,4$ are additionally
constrained by the condition $\sum_{l=1}^4 x_{2l} =0$.
For the groups $\m Z_7 \rtimes \m Z_3$, $\m A_4$, $\m D_5$, and $\m S_3$ there
are two inequivalent ways to assign discrete quadratic indices. Therefore,
anomaly freedom of the underlying continuous family symmetry $G_f$ requires
that the anomaly conditions be satisfied for {\it both} of these choices
separately.  When calculating the discrete anomaly $\m G - \m G - U(1)_Y$ it
is necessary to choose the normalization with $Y_Q=1$.

The first table shows the finite groups which are subgroups of $SU(3)$
only. Therefore the discrete cubic anomaly provides a useful condition. In the
second table we list the finite groups which can be considered as subgroups of
$SO(3)$ as well. Hence, the discrete cubic anomaly condition of $\m A_4
\subset SU(3)$ is omitted, see Section~\ref{DACsect}.

\vspace{1mm}

{\small
\begin{center}
$
\begin{array}{||c||rr|rrr|rr||}\hline\hline &&&&&&& \\[-3mm]
\m G \subset SU(3) & \multicolumn{2}{|c|}{\m{PSL}_2(7)} & \multicolumn{3}{|c|}{\m Z_7 \rtimes \m Z_3} & \multicolumn{2}{|c||}{\Delta(27)} \\[2mm] \hline\hline &&&&&&& \\[-3.5mm]
& ~{\bf 3}: & ~~1~~ & ~{\bf 1'}:~ & ~~1/2~~ & ~~2~~ & ~{\bf 1_{2l}}: & ~~x_{2l}~~ \\[1.5mm]
& ~\ol{\bf 3}: & ~~1~~ & ~\ol{\bf 1'}:~ & ~~1/2~~ & ~~2~~ & ~{\bf 1_{2l-1}}: & ~~x_{2l}~~ \\[1mm]
~~\wt \ell({\bf r_i})~~& ~{\bf 6}: & ~~5~~ & ~{\bf 3}:~ & ~~1~~ & ~~1~~ & ~{\bf 3}: & ~~1~~ \\[1.5mm]
& ~{\bf 7}: & ~~14~~ & ~\ol{\bf 3}:~ & ~~1~~ & ~~1~~ & ~\ol{\bf 3}: & ~~1~~ \\[1.5mm]
& ~{\bf 8}: & ~~6~~ &  & &  &  &  \\[1.5mm] \hline &&&&&&& \\[-4mm]
\sum_{i=\mathrm{light}} \wt\ell_iY_i & \multicolumn{2}{|c|}{0\mathrm{~mod~}24} & \multicolumn{3}{|c|}{0 \mathrm{~mod~}3} &\multicolumn{2}{|c||}{0 \mathrm{~mod~}3} \\[2mm] \hline\hline &&&&&&& \\[-3.5mm]
& ~{\bf 3}: & ~~1~~ & ~{\bf 1'}:~ & ~~y & ~~~~ & ~{\bf 1_{2l}}: & ~~y_{2l}~~ \\[1.5mm]
& ~\ol{\bf 3}: & ~~-1~~ & ~\ol{\bf 1'}:~ & ~~-y & ~~~~ & ~{\bf 1_{2l-1}}: & ~~-y_{2l}~~ \\[1mm]
~~\wt A({\bf r_i})~~& ~{\bf 6}: & ~~0~~ & ~{\bf 3}:~ & ~~1 & ~~~~ & ~{\bf 3}: & ~~1~~ \\[1.5mm]
& ~{\bf 7}: & ~~0~~ & ~\ol{\bf 3}:~ & ~~-1 & ~~~~ & ~\ol{\bf 3}: & ~~-1~~ \\[1.5mm]
& ~{\bf 8}: & ~~0~~ &  & &  &  &  \\[1.5mm] \hline &&&&&&& \\[-4mm]
\sum_{i=\mathrm{light}} \wt A_i & \multicolumn{2}{|c|}{0  \mathrm{~mod~}7} &\multicolumn{3}{|c|}{0 \mathrm{~mod~}7} &\multicolumn{2}{|c||}{0\mathrm{~mod~}9} \\[2mm] \hline\hline
\end{array}
$
\end{center}

\vspace{-0.5mm}

\begin{center}
$
\begin{array}{||c||rr|rrr|rrr|rrr||}\hline\hline &&&&&&&&&&& \\[-3mm]
\m G \subset SO(3) & \multicolumn{2}{|c|}{\m{S}_4} & \multicolumn{3}{|c|}{\m
  A_4} & \multicolumn{3}{|c|}{\m D_5} & \multicolumn{3}{|c||}{\m S_3}\\[2mm] \hline\hline &&&&&&&&&&& \\[-3.5mm]
& {\bf 1'}: & ~13-x & {\bf 1'}: & ~2~ & ~8 & {\bf 1'}: & ~5/2~ & ~25/2&{\bf 1'}: & -3/2~ & ~9/2\\[1.5mm]
\wt\ell({\bf r_i})& {\bf 2}\,: & ~5-x & \ol{\bf 1'}: & ~2~ & ~8 & {\bf 2_1}: & -3/2~ & ~17/2&{\bf 2}\,: & ~5/2~ & ~17/2\\[1.5mm]
& {\bf 3_1}: & ~x & {\bf 3}: & ~1~ & ~1 & {\bf 2_2}: & ~13/2~ & ~33/2& &  & \\[1.5mm]
& {\bf 3_2}: & ~1 &  &  & &  &  & &  &  & \\[1.5mm] \hline  &&&&&&&&&&&   \\[-4mm]
\sum_{i=\mathrm{light}} \wt\ell_iY_i & \multicolumn{2}{|c|}{0\mathrm{~mod~}24} & \multicolumn{3}{|c|}{0 \mathrm{~mod~}12} &\multicolumn{3}{|c|}{0 \mathrm{~mod~}20} &\multicolumn{3}{|c||}{0 \mathrm{~mod~}12}\\[2mm] \hline\hline 

\end{array}
$
\end{center}
}

\end{appendix}




\end{document}